\documentclass[prodmode,acmtaas]{acmsmall}

\usepackage{url}
\usepackage{wrapfig}
\usepackage[caption=false,font=footnotesize]{subfig}

\newcommand{\elem}[1]{\textsf{\small #1}}

\usepackage{xspace}
\newcommand{\EUREMA}{{\mbox{EUREMA}}\xspace}
\newcommand{\FeedbackLoopDiagram}{{FLD}\xspace}
\newcommand{\FeedbackLoopDiagrams}{{FLDs}\xspace}
\newcommand{\LayerDiagram}{{LD}\xspace}
\newcommand{\LayerDiagrams}{{LDs}\xspace}
\newcommand{\mRUBiS}{{\mbox{mRUBiS}}\xspace}

\usepackage[hidelinks]{hyperref}
\hypersetup{
	bookmarks=true,         
	unicode=false,          
	pdftoolbar=true,        
	pdfmenubar=true,        
	pdffitwindow=false,     
	pdftitle={Model-Driven Engineering of Self-Adaptive Software with EUREMA},
	pdfauthor={Thomas Vogel and Holger Giese},
	pdfsubject={},
	pdfcreator={},
	pdfproducer={},
	pdfkeywords={model-driven engineering, modeling language, models at runtime, model interpreter, self-adaptive software, feedback loops, layered architecture, software evolution}, 
	pdfnewwindow=true,
}

\acmVolume{8}
\acmNumber{4}
\acmArticle{18}
\acmYear{2014}
\acmMonth{01}
\doi{2555612}

\begin{document}
\markboth{T. Vogel and H. Giese}{Model-Driven Engineering of Self-Adaptive Software with EUREMA}
\title{Model-Driven Engineering of Self-Adaptive Software with EUREMA}
\author{THOMAS VOGEL and HOLGER GIESE
\affil{Hasso Plattner Institute, University of Potsdam}}

\begin{abstract}
The development of self-adaptive software requires the engineering of an adaptation engine that controls the underlying adaptable software by feedback loops. The engine often describes the adaptation by runtime models representing the adaptable software and by activities such as analysis and planning that use these models. 
To systematically address the interplay between runtime models and adaptation activities, \emph{runtime megamodels} have been proposed. A runtime megamodel is a specific model capturing runtime models and adaptation activities.
In this article, we go one step further and present an executable modeling language for \emph{ExecUtable RuntimE MegAmodels} (\EUREMA) that eases the development of adaptation engines by following a model-driven engineering approach. We provide a domain-specific modeling language and a runtime interpreter for adaptation engines, in particular feedback loops. Megamodels are kept alive at runtime and by interpreting them, they are directly executed to run feedback loops. Additionally, they can be dynamically adjusted to adapt feedback loops. Thus, EUREMA supports development by making feedback loops explicit at a higher level of abstraction and it enables solutions where multiple feedback loops interact or operate on top of each other and self-adaptation co-exists with off-line adaptation for evolution.
\end{abstract}

\category{D.2.2}{Software Engineering}{Design Tools and Techniques}
\category{D.2.9}{Software Engineering}{Management}
\category{D.2.10}{Software Engineering}{Design}

\terms{Design, Languages}
\keywords{Model-driven engineering, modeling language, models at runtime, model interpreter, self-adaptive software, feedback loops, layered architecture, software evolution}
\acmformat{Vogel,\,T. and Giese,\,H. 2014. Model-driven engineering of self-adaptive software with EUREMA.}
\begin{bottomstuff}
Authors' addresses: T. Vogel (corresponding author) {and} H. Giese, Hasso Plattner Institute for Software Systems Engineering, University of Potsdam, Germany; email: thomas.vogel@hpi.uni-potsdam.de
\end{bottomstuff}

\maketitle


\section{Introduction}\label{sec:introduction}

Self-adaptation capabilities are required for many modern software systems that are self-aware, context-aware, mission-critical, or ultra-large-scale to dynamically adapt their configuration in response to changes in the system itself, the environment, or the requirements \cite{SEfSAS-ROADMAP-2009,SEfSAS-ROADMAP-2012}.
The development of self-adaptive software following the \emph{external approach}~\cite{Salehie&Tahvildari2009} separates the software into the \emph{adaptable software} and the \emph{adaptation engine}. In between both, a \emph{feedback loop} ensures that the adaptation engine dynamically adjusts the adaptable software if needed.
This separation decouples the engine from the adaptable software but it makes the feedback loop a crucial element of the overall software architecture \cite{Shaw95beyondobjects,PezzeMS08,Brun+2009} and therefore essential for engineering an adaptation engine in the external approach. 

In this article, we present a model-driven engineering (MDE) approach called \emph{\mbox{ExecUtable} \mbox{RuntimE} \mbox{MegAmodels}} (\EUREMA) that enables the specification and execution of adaptation engines for self-adaptive software with multiple feedback loops. 
The \EUREMA language eases the development of adaptation engines by supporting a domain-specific modeling solution and the \EUREMA runtime interpreter supports the execution of adaptation engines and feedback loops.
Moreover, \EUREMA explicitly maintains the runtime models used within an adaptation engine, the interplay between these models, and the adaptation activities working on these models. Thus, the maintenance and evolution of runtime models and adaptation activities continues at runtime beyond the initial development of the software.

The \EUREMA modeling language is specific for adaptation engines and it provides two types of diagrams to specify them.
A behavioral \emph{feedback loop diagram} (\FeedbackLoopDiagram) is used to model a feedback loop or individual adaptation activities and runtime \mbox{models} of a loop. An \FeedbackLoopDiagram is considered as a \emph{megamodel module} specification encapsulating the details of a partial or complete feedback loop.
A structural \emph{layer diagram} (\LayerDiagram) describes how the megamodel modules and the adaptable software are related to each other in an instance situation of the self-adaptive software.
Thus, an \LayerDiagram provides an architectural view that considers feedback loops encapsulated in modules as black boxes while white-box views are provided by \FeedbackLoopDiagrams.
Hence, \EUREMA models specify feedback loops and their structuring in adaptation engines. These models make the feedback loops explicit in the architectural design of self-adaptive software and they are kept alive at runtime and executed by an interpreter. This supports the seamless design, development, execution, and even adaptation of feedback loops.

This article discusses \EUREMA with its following contributions:
(a)~we thoroughly discuss requirements for adaptation engines, 
(b)~feedback loops and their coordinated execution are explicitly modeled,
(c)~a feedback loop's knowledge is explicitly captured by runtime models,
(d)~\EUREMA models are kept alive at runtime and they are directly executed by an interpreter, which leverages adaptive feedback loops in layered architectures,
(e)~self-adaptation and off-line adaptation co-exist for evolving self-adaptive software,
and 
(f)~we evaluate \EUREMA by modeling state-of-the-art \mbox{approaches} to self-adaptive software from literature and by quantifying the runtime efficiency of the interpreter.

This article is a revised and extended version of~\cite{VG12b} that introduced the initial concepts of \EUREMA and therefore only addressed the contributions~(b),~(c), and~(d). In contrast, this article refines them by extending the language with layer diagrams (\LayerDiagrams) including feedback loop triggers, and by proposing a sound approach for layered architectures. Therefore, this article entirely presents the novel contributions (a),~(e), and~(f).

The rest of the article is structured as follows. Foundations and requirements for self-adaptive software are discussed in Section~\ref{sec:requirements} and related work in Section~\ref{sec:state-of-the-art}. We introduce the \EUREMA concepts as proposed by \cite{VG12b} in Section~\ref{sec:seams2012} and discuss the novel concepts in Section~\ref{sec:ld}. We present technical details in Section~\ref{sec:implementation}, discuss design decisions and the requirements coverage in Section~\ref{sec:discussion}, and evaluate \EUREMA in Section~\ref{sec:evaluation}. Finally, we conclude the article and outline future work.

\section{Terminology, Concepts, and Requirements}\label{sec:requirements}

\begin{figure}[t]
\begin{minipage}[b]{0.31\linewidth}
\centerline{\includegraphics[height=28.5mm]{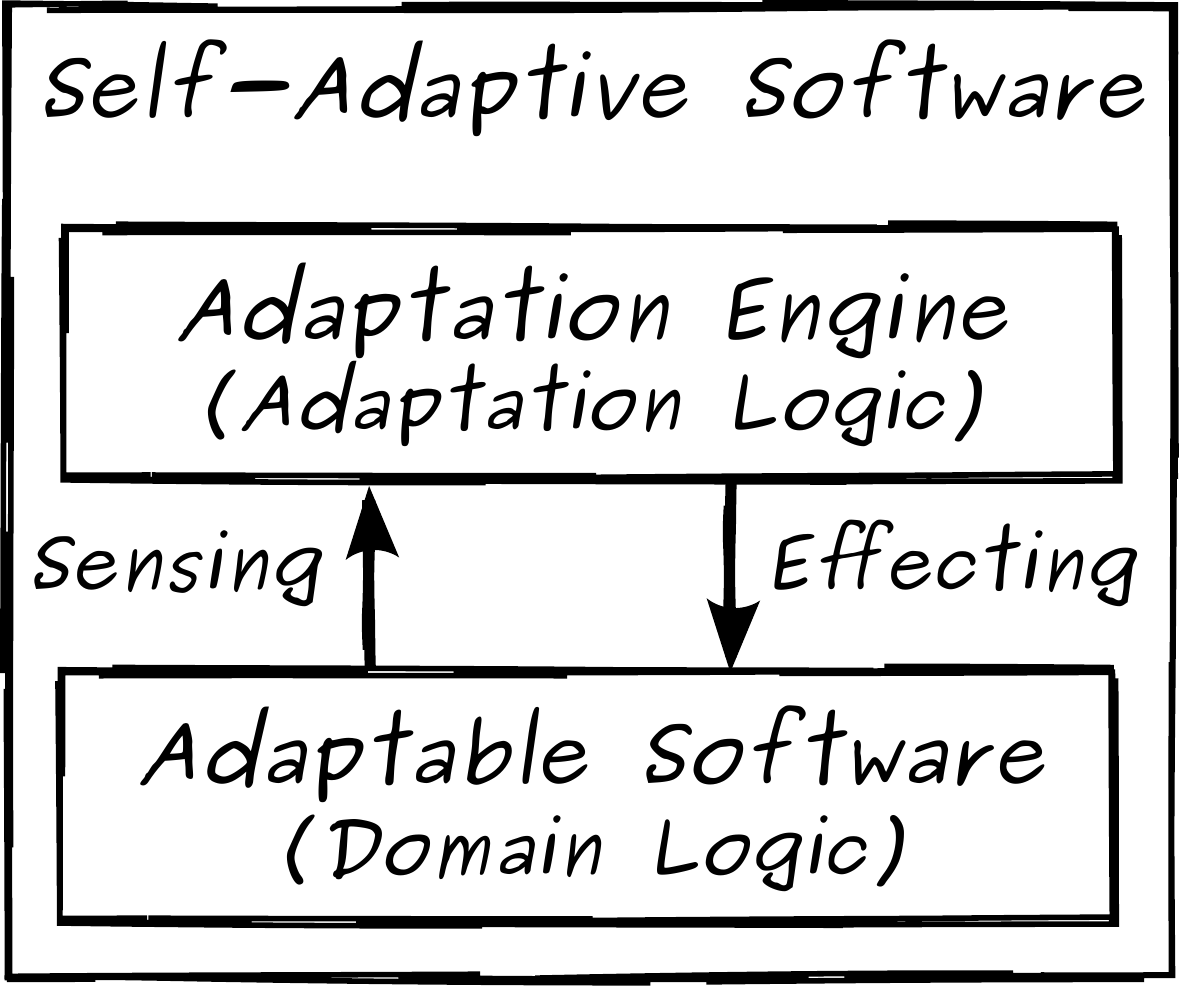}}
\caption{External approach}
\label{fig:external-approach}
\end{minipage}%
\begin{minipage}[b]{0.31\linewidth}
\centerline{\includegraphics[height=28.5mm]{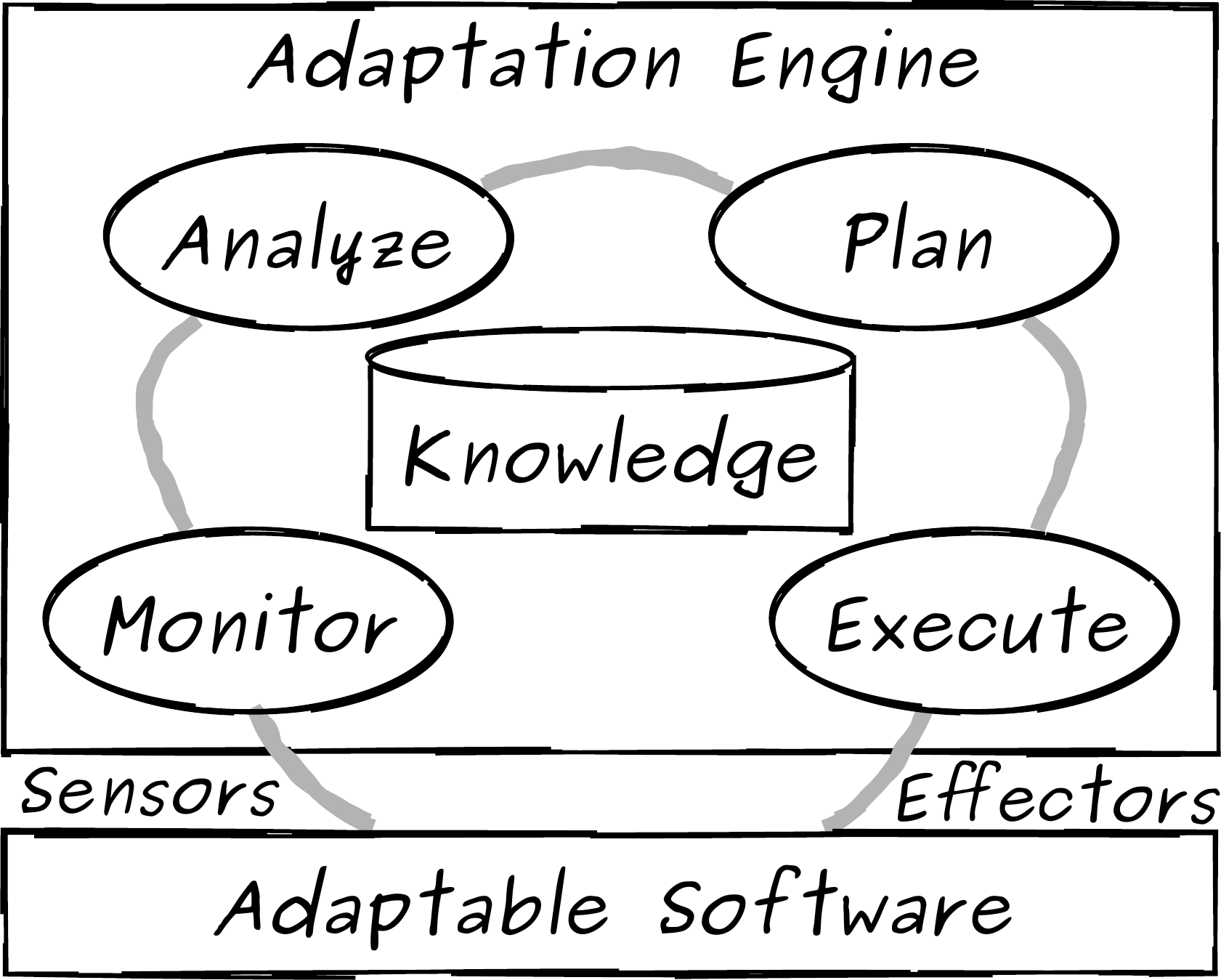}}
\caption{MAPE-K}
\label{fig:MAPE-K}
\end{minipage}%
\begin{minipage}[b]{0.37\linewidth}
\centerline{\includegraphics[height=28.5mm]{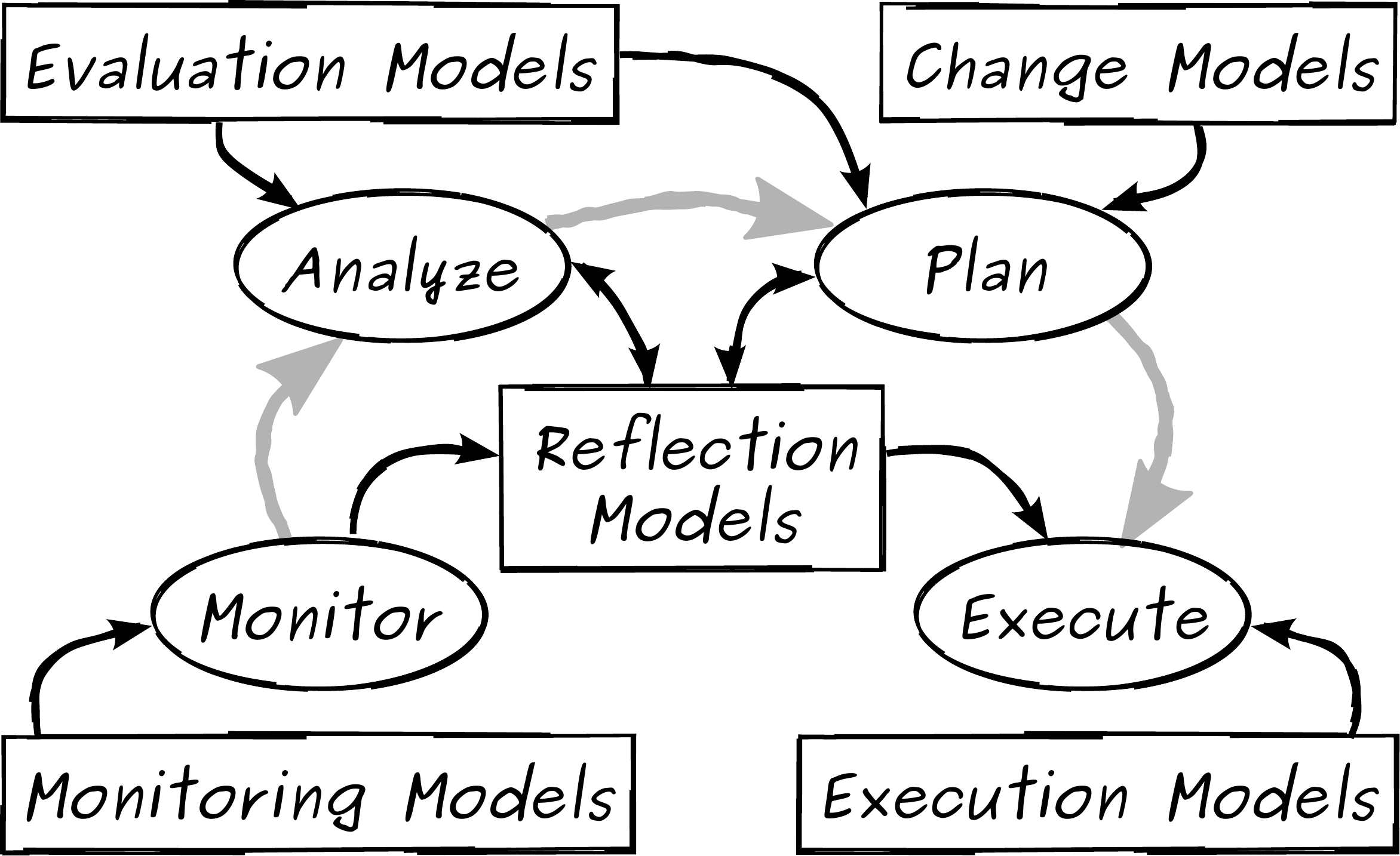}}
\caption{Runtime Models}
\label{fig:MAPE+Models}
\end{minipage}
\end{figure}

In this section, we clarify relevant terminology and concepts of self-adaptive software and we discuss core requirements for engineering such software.
As the \emph{external approach} is typically adopted in self-adaptive software \cite{Salehie&Tahvildari2009}, we consider this approach as depicted in Figure~\ref{fig:external-approach}. It assumes a basic architecture that splits the \emph{self-adaptive software} into the \emph{adaptation engine} and the \emph{adaptable software} while the former one controls (\emph{sensing} and \emph{effecting}) the latter one. The adaptable software realizes the \emph{domain logic} and the engine implements the \emph{adaptation logic} as a~feedback loop, which constitutes self-adaptation.

Thus, the engineering of adaptation engines and feedback loops is essential for the external approach. This requires a modeling language and techniques to design, implement, run, and maintain such an engine with its feedback loops.   
In the following, we discuss corresponding requirements (\textbf{R}) for self-adaptive software, particularly for a modeling language.

\paragraph{Feedback Loops}
Separating the adaptation engine from the adaptable software makes the \emph{feedback loop} between them a crucial element of the architecture, which has to be made explicit in the design of self-adaptive software \cite{Brun+2009,PezzeMS08,Shaw95beyondobjects}. Thus, feedback loops have to be explicitly modeled (\textbf{R1}).
A more detailed view of the feedback loop is provided by the \emph{MAPE-K} cycle (Monitor/Analyze/Plan/Execute-Knowledge)~\,\cite{Kephart&Chess2003}~\,depicted in Figure~\ref{fig:MAPE-K}. The feedback loop is refined to four adaptation activities sharing  knowledge. The adaptable software is \emph{monitored} and \emph{analyzed}, and if changes are required, adaptation is \emph{planned} and \emph{executed} to this software.

As sketched in Figure~\ref{fig:MAPE-K}, the modeling language should support the specification of adaptation activities that form a feedback loop. This includes the \emph{intra-loop coordination} \cite{Vromant+2011} (\textbf{R2}) by means of the control flow for these activities, which makes the execution dependencies between individual activities explicit.
Moreover, the language should address \emph{when} a feedback loop should be executed, for example, by capturing \emph{triggering conditions} (\textbf{R3}).

Additionally, multiple feedback loops have to be considered \cite{Kephart&Chess2003,Weyns+2012} to handle different concerns such as failures and performance \cite{Kephart+2007,VG10} or to decentralize control \cite{SEfSAS2-decentral}.
The language should therefore support the modeling of multiple, interacting feedback loops. This imposes the need for \emph{inter-loop coordination} \cite{Vromant+2011} (\textbf{R4}) and \emph{distribution} (\textbf{R5}) of feedback loops.
Moreover, besides specifying feedback loops, the language should support the \emph{concurrent execution} (\textbf{R6}) of the feedback loops and activities based on the specifications.

\paragraph{Knowledge}
In MAPE-K, the adaptation activities are the computations performing self-adaptation and the knowledge refers to data used by these computations. This motivates the explicit treatment of the \emph{knowledge} that is refined in our case to a set of \emph{runtime models}. While \citeN{MC.2009.326} consider only runtime models that reflect the adaptable software, we take a broader perspective and consider all models that are used by any adaptation activity. Thus, analysis rules and adaptation strategies are examples for further runtime models. Based on a literature review, we proposed a categorization of runtime models for feedback loops in~\cite{VogelSG11}, which is depicted in an extended version in Figure~\ref{fig:MAPE+Models}.

\emph{Reflection Models} reflect the adaptable software and its environment, and they are~updated by monitoring the software and environment. Thereby, \emph{Monitoring \mbox{Models}} map system-level observations to the abstraction level of reflection models. The reflection models are analyzed to identify adaptation needs by applying \emph{Evaluation Models} that, for example, define constraints on reflection models. If adaptation needs have been identified, the planning activity devises a plan prescribing the adaptation on the reflection models. Planning is specified by \emph{Change Models} describing the adaptable software's variability space. Evaluation models such as utility preferences guide the exploration of this space to find an appropriate adaptation. Finally, the execute activity enacts the planned adaptation on the adaptable software based on \emph{Execution Models} that refine model-level adaptation to system-level adaptation.

Evaluation and change models do not have to be strictly separate models, which is exemplified by event-condition-action rules that address the analyze (evaluating the condition) and the plan (applying the actions) activities in one step. Thus, we combine them to \emph{Adaptation Models} \cite{VG12}. Monitoring and execution models are concerned with the synchronization of the adaptable software and the reflection models. This is known as the causal connection~\cite{Maes1987} such that we consider them as \emph{Causal Connection Models}.

This categorization shows that different kinds of runtime models are simultaneously used in a feedback loop, which should be captured by a modeling language for such loops (\textbf{R7}).

\paragraph{Sensors and Effectors}
The adaptation engine and the adaptable software are connected by sensors and effectors (cf.~Figure~\ref{fig:MAPE-K}). Thus, the modeling language has to cover when the monitor and execute activities that use the sensors and effectors are activated to maintain the causal connection between the reflection models and the adaptable software.
Since sensors and effectors usually depend on the specific adaptable software, we---and others like \citeN{GarCHSS04} and \citeN{1555028}---assume that they are provided by the software and that most of their details are hidden in the implementation of the monitor and execute activities. This assumption is motivated by programming language and middleware platforms that have recognized the need for runtime management by supporting the development of sensors and effectors or already providing them through application programming interfaces. Examples are the \emph{Java Management Extensions (JMX)} or the \emph{OSGi} platform.

Depending on the available sensors and effectors, \emph{parameter} or \emph{structural adaptation}, or a combination of both \cite{McKinley+2004} can be realized. Parameter adaptation changes variables and structural adaptation the architecture of the adaptable software. Focusing on software architectures, we especially require support for structural adaptation (\textbf{R8}).

\paragraph{Layered Architecture}
Feedback loops might have to operate on top of each other, which results in layers of feedback loops, in order to realize adaptive~\cite{Isermann+1992,Kokar+1999} or hierarchical control schemes \cite{Findeisen1980,Hestermeyer+2004}, robot software \cite{Gat1997}, or the reference architecture of \citeN{Kramer&Magee2007}. 

In such layered architectures, a feedback loop at a higher layer can adapt the feedback loop at the layer below. Therefore, some form of reflection of the lower-layer feedback loop has to be provided at runtime enabling the adaptation of this loop~\cite{Andersson+2009}. 

Thus, the modeling language should leverage layered architectures by supporting adaptable feedback loops and reflection models representing these loops for structural adaptation (\textbf{R9}). In this context, \emph{declarative} and \emph{procedural reflection} \cite{Maes1987} should be supported. In declarative reflection, a separate representation of the program is maintained and used for meta-computations such as adaptation. Procedural reflection maintains no separate representation and uses directly the program for meta-computations. 
Finally, the language should support feedback loops that adapt lower-layer loops by operating on the reflection models.

\paragraph{Offline Adaptation}
Though self-adaptation promises that the software adjusts itself by automating adaptation activities otherwise performed offline for evolution, we cannot expect that software is able to cope with \emph{all} needs for evolution and to fully automate \emph{all} kinds of offline activities.
Thus, besides an adaptation engine realizing (online) self-adaptation, the engine's co-existence with offline adaptation is required \cite{GGH08_ag,SEfSAS2-process}.
Similar to~\citeN{SEfSAS2-process}, we consider an adaptation activity to be \emph{offline} if it is performed externally to the self-adaptive software as typically done today in development environments. If an \mbox{activity} is performed internally to the software, we refer to \emph{online} activities.
Hence, the language and its runtime environment should support the co-existence of online and \mbox{offline} adaptation for evolving self-adaptive software (\textbf{R10}).

\vspace{.25em}
\section{State of the Art in Engineering Adaptation Engines}\label{sec:state-of-the-art}

\vspace{.5em}

In the following, we review state-of-the-art approaches for engineering adaptation engines with respect to the requirements discussed in the previous section.

There exists a lot of work on feedback loops to control systems, like in \emph{autonomic computing} that applies control theory to parameter adaptation of software~\cite{Hellerstein+2004,Kokar+1999}. However, self-adaptation oftentimes considers dynamic software architectures \cite{McKinley+2004} (\textbf{R8}), which prevents a direct application of control theory and requires new means for engineering.
Popular means are frameworks that use some form of models \cite{Salehie&Tahvildari2009}.
Such frameworks employ models to specify self-adaptation as mappings of assertions to adaptation actions \cite{SchmWhiGok07} or as transitions between configurations of the adaptable software \cite{BB09}. These models are used for generating partial adaptation engines to ease development. However, the resulting engines supporting single feedback loops are structurally static and pre-defined by the frameworks. The models do not make the feedback loop explicit (\textbf{R1}) and they are not kept alive at runtime (\textbf{R7}) to execute (\textbf{R6}) or dynamically adjust (\textbf{R9}) the engine.

In contrast, frameworks such as 
\emph{Rainbow} \cite{GarCHSS04}, 
\emph{MADAM} \cite{1128711}, 
\emph{MUSIC} \cite{RouvoyEtAl09}, 
\emph{DiVA} \cite{Morin+2009,1555028}, or 
\emph{GRAF} \cite{Amoui20122720} 
maintain runtime models that specify the adaptation and capture the feedback loop's knowledge (\textbf{R7}). These models can be modified at runtime by engineers, especially to replace adaptation strategies to adjust the adaptation logic. However, support for dynamically adjusting a feedback loop (\textbf{R9}) is limited since these frameworks support only single loops, whose structuring of adaptation activities cannot be adjusted in contrast to specific models consumed by the activities. Additionally, the runtime models do not explicitly specify feedback loops (\textbf{R1}) because these frameworks prescribe single feedback loops and just offer customization points, like to inject adaptation strategies. This is motivated by their focus to reduce efforts for developing adaptation engines at the expense of limited flexibility. Thus, when developing a specific self-adaptive software, these frameworks do not support feedback loops that are entirely and individually designed by engineers for the specific case.

All the approaches discussed so far do not support adaptation engines with multiple feedback loops (\textbf{R4}).
\citeN{Kephart+2007} consider interactions between two feedback loops that manage competing concerns by a solution ``established through trial and error''~\cite[p.~24]{Kephart+2007} for one specific case. 
In contrast, a generic coordi\-nation protocol is presented by \citeN{Oliveira+2012}, which supports mutual exclu\-sive access to knowledge and the triggering among the loops. However, the protocol~is restricted as a loop may only trigger another loop from the execute but not from the monitor, analyze, or plan activities. Thus, directly coordinating, for example, the individual analyses is not supported. 
In \cite{Gueye+2012}, the coordination between feedback loops is realized by a controller that decides which loop may exclusively operate. This decision is specified by coordination models and policies used for generating the controller. However, these models do not specify the feedback loops and their coordination at the architectural level of self-adaptive software (\textbf{R1}).
Other approaches addressing multiple loops are implementation frameworks that aim at reducing development efforts without prescribing a specific coordination mechanism. \citeN{Vromant+2011} provide reusable components for the distributed communication among feedback loops or adaptation activities. \citeN{ChengWICSA2004} provide an abstraction layer between the adaptable software and the feedback loops, through which all loops have consistent access and knowledge about the software. However, this layer does not \mbox{coordinate}~the~feedback loops. 

All of the approaches discussed so far provide specific and pre-defined solutions or generic implementation support, which results in feedback loops whose structure cannot be dynamically adapted.
Such adaptation is addressed by layered architectures, in which a higher-layer feedback loop adjusts the feedback loop at the layer below (\textbf{R9}). 
In our previous work on \emph{Mechatronic UML}, we extended UML to specify and generate a hierarchical scheme that addresses control, reconfiguration, and planning by distinct feedback loops at different layers~\cite{Hestermeyer+2004}. However, the adaptation is defined before deployment and cannot be dynamically changed, especially since the models are not kept alive at runtime (\textbf{R7}). 
\citeN{Heaven2009} propose a similar three-layer architecture, in which
plans generated by the highest layer are executed by the middle layer to generate new configurations for the lowest layer. Thus, a layer adjusts another layer by providing new plans or configurations. However, the solution focuses on synthesizing initial plans before deployment but it does not support replanning at runtime and thus the highest-layer feedback loop adapting the middle layer.
In contrast, \emph{PLASMA}~\cite{Tajalli2010} supports adapting the middle layer in a similar architecture. However, the extent of this adaptation is not clear since the middle layer architecture is predefined by engineers. The focus of \emph{PLASMA} is to automate plan generation and enactment while the employed feedback loops, adaptation activities, and knowledge are not explicitly modeled for all layers (\textbf{R1}). 
Finally, the numbers of layers (three) and feedback loops for each layer (one) seem to be immutable. Thus, multiple feedback loops for a layer, or (dynamically) changing the number of feedback loops and layers are not supported.

Such extensive changes can be seen as an evolution of the self-adaptive software performed through offline adaptation (\textbf{R10}). \citeN{GGH08_ag} discuss the idea of having two intertwined feedback loops for self-adaptation and offline adaptation but they present no working solution.
As discussed previously, frameworks utilizing runtime models often support changing those models, for example, to replace adaptation strategies at runtime. \citeN{Morin+2009} claim to support evolution as changes performed manually on runtime models in the development environment. They propose an initial solution in \cite{MorinCIT2009}, which, however, does not consider changes in the structure or number of feedback loops or layers. 

All approaches discussed so far do not make the feedback loops, their adaptation activities, runtime models, and coordination explicit in the architectural design (\textbf{R1}).~In this context, only a few approaches exist.
We \cite{HGB2010} proposed a UML \mbox{profile} to make feedback loops and the interplay of multiple loops explicit in the design, however, without capturing the loops' individual adaptation activities (\textbf{R2}) and runtime models (\textbf{R7}).
In contrast, \citeN{Weyns+2012} present a formal reference model that captures feedback loops including the activities and models. The goal of the reference model is to support engineering by studying early design alternatives. 
However, the models created by both approaches are used for the architectural design but they are not kept alive at runtime, for example, for execution~(\textbf{R6}).

Summing up, state-of-the-art approaches for engineering self-adaptive software aim at reducing development efforts by generating adaptation engines or providing frameworks. The resulting engines often consist of single feedback loops whose structure is rather static and predefined. This limits their adaptation during development, dynamically in layered architectures, and offline by engineers. In general, there exist only preliminary work on layered architectures (\textbf{R9}) and offline adaptation (\textbf{R10}).
Moreover, approaches providing runtime support for self-adaptation (\textbf{R6}) do not address the explicit modeling of feedback loops (\textbf{R1}). Thus, they do not consider runtime models such as executable megamodels that describe feedback loops and leverage the adaptation of feedback loops.
In contrast, approaches tackling the explicit modeling (\textbf{R1}) are focused on the design and they do not provide any runtime support based on these models (\textbf{R6}).
Thus, state-of-the-art approaches do not support all the requirements for engineering adaptation engines discussed in Section~\ref{sec:requirements}.

In contrast, we propose a seamless approach whose modeling language supports the explicit design (\textbf{R1}), execution (\textbf{R6}), and adaptation (\textbf{R9, R10}) of feedback loops. 
Thereby, \EUREMA improves the state of the art concerning frameworks because it does not prescribe any structure of the adaptation activities or feedback loops and it does not limit the number of feedback loops or layers.
In contrast to existing modeling languages, \EUREMA provides improvements by keeping the models alive at runtime for executing feedback loops and for adjusting feedback loops and layers either dynamically or by offline adaptation.

\section{Feedback Loop Diagrams}\label{sec:seams2012}
This section discusses the \EUREMA language for engineering adaptation engines as introduced in \cite{VG12b}. The language is based on the concept of megamodels originating from the research field of model management in MDE. A \emph{megamodel} refers to a model that contains other models and relationships between these models while the relationships constitute operations such as model transformations \cite{BDDFB07,Bezivin_et_al:2003,Bezivin+2004,favre:DSP:2005:13}. 
\EUREMA adopts this generic concept for specifying and executing feedback loops by considering the feedback loop's knowledge as runtime models and the individual adaptation activities as model operations working these runtime models.

Therefore, \EUREMA (mega)models explicitly specify feedback loops by capturing the runtime models, the interplay and usage of these models, and the flow of model operations. Moreover, \EUREMA models are kept alive at runtime for executing feedback loops. Thereby, megamodel concepts are leveraged at runtime to explicitly maintain and evolve the runtime models and model operations beyond the initial development of the feedback loops.

To discuss \EUREMA and all of its modeling concepts, we keep the adaptable software abstract to cover different variants of adaptation engines instead of a specific engine for a specific software. 
We consider a component-based application as the adaptable software, which is adaptable at the architectural level. Tackling adaptation at the architectural level is a popular approach \cite{GarCHSS04,1555028} as it provides promising abstractions for parameter and structural adaptation \cite{McKinley+2004}.
Nevertheless, most examples in this article are extended and generalized scenarios of our earlier work \cite{VG10,VogelNHGB10} that used an online marketplace, recently extended to \emph{\mRUBiS}\cite{mRUBiS}, as an application example. However, \EUREMA has not been used in our earlier work that just proposed a code-based and static solution for structuring the adaptation activities and runtime models of the feedback loops.
In contrast, \EUREMA proposes a model-driven and flexible approach to specify, execute, and adapt feedback loops.

\subsection{Modeling an Individual Feedback Loop}\label{sec:seams2012:FLD}
\begin{figure}[t]
\hspace{0.01\linewidth}
\begin{minipage}[b]{0.58\linewidth}
\centerline{\includegraphics[scale=.264]{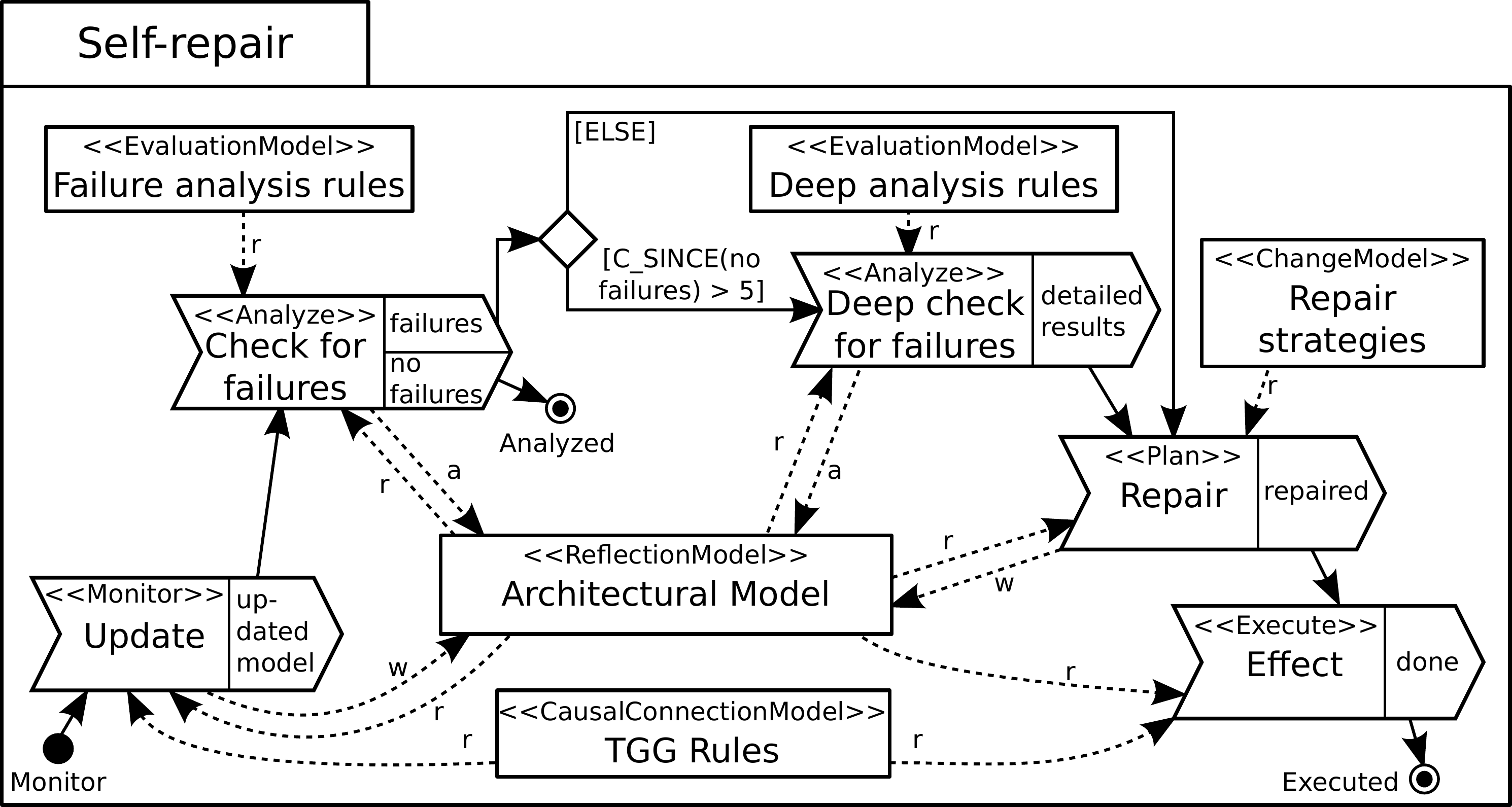}}
\caption{\FeedbackLoopDiagram for {\sf Self-repair}}
\label{fig:mm:self-repair}
\end{minipage}%
\hspace{0.01\linewidth}
\begin{minipage}[b]{0.36\linewidth}
\centerline{\includegraphics[scale=.264]{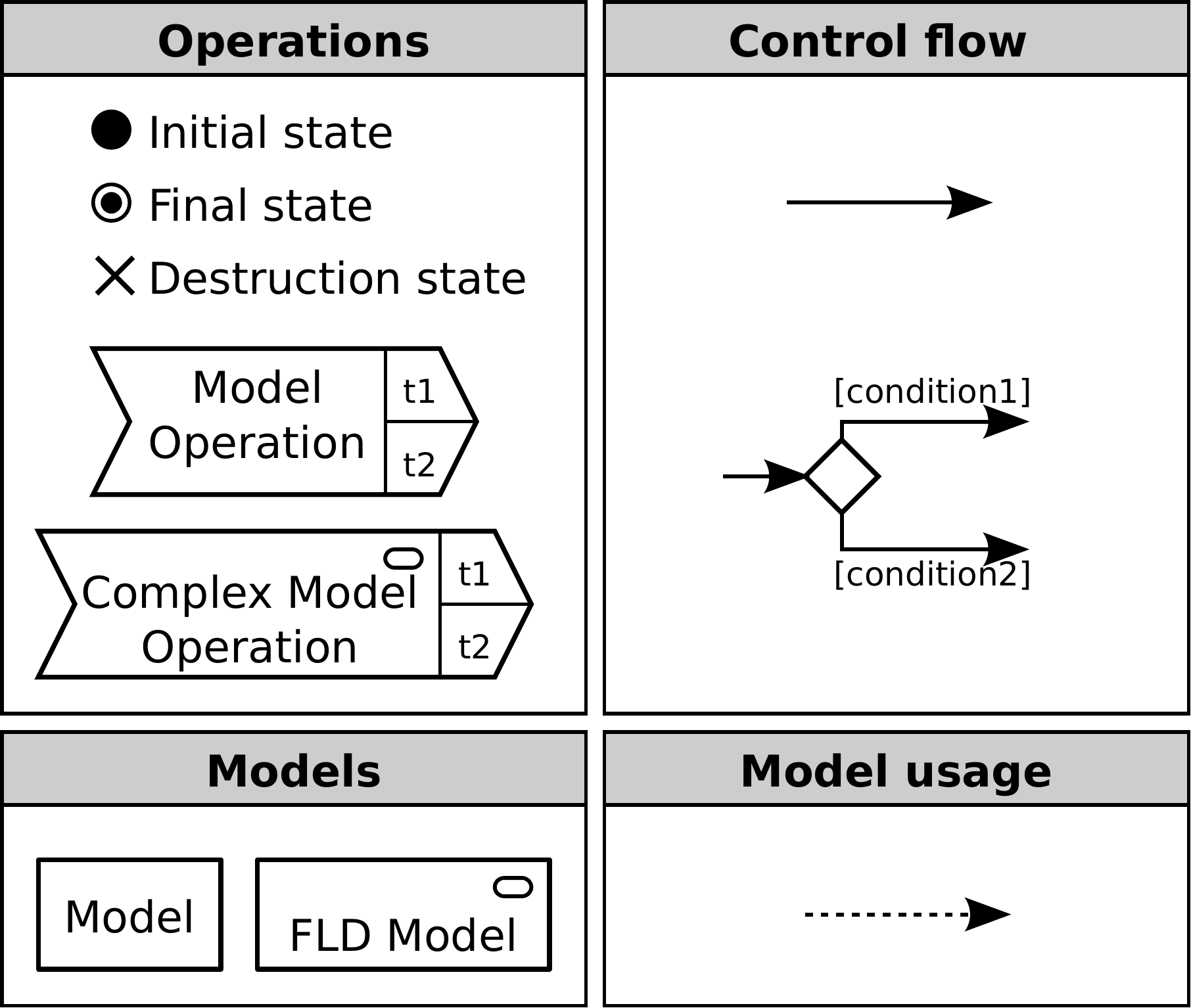}}
\caption{Concrete syntax for \FeedbackLoopDiagrams}
\label{fig:concrete-syntax-FLD}
\end{minipage}
\hspace{0.01\linewidth}
\end{figure}

To model a feedback loop with its adaptation activities and runtime models, the \EUREMA language provides a behavioral \emph{feedback loop diagram} (\FeedbackLoopDiagram). Such a diagram specifies a feedback loop by means of \emph{operations}, the \emph{control flow} of operations, runtime \emph{models}, and the \emph{model usage} by operations.
For instance, a feedback loop supporting self-repair capabilities by automatically recovering the adaptable software from failures at runtime is specified by the \FeedbackLoopDiagram depicted in Figure~\ref{fig:mm:self-repair}. This \FeedbackLoopDiagram is framed and labeled with its name \elem{Self-repair}.

The concrete syntax of \FeedbackLoopDiagrams is shown in Figure~\ref{fig:concrete-syntax-FLD}.
\emph{Initial} and \emph{final states} are \mbox{special} operations that define the entry and exit points for executing a feedback loop~\mbox{instance}. A \emph{destruction state} is a final state that destroys the instance when this state is reached. 
Adaptation activities are specified as \emph{model operations} represented by hexagon block arrows labeled with their names. A model operation has at least one named exit \mbox{compartment}, one for each return status of the operation. At runtime, the implemen\-tation of the operation determines the return status and therefore, which compart\-ment~is activated to continue the control flow. \emph{Complex model operations} abstract~from and invoke adaptation activities modeled in other \FeedbackLoopDiagrams, which will be discussed in Section~\ref{sec:seams2012:modularFLD}.
The \emph{control flow} between operations is specified by solid arrows and can be exclusively branched by a decision node (diamond element) and conditions. The language for these conditions refers to counter and timing information about the execution of the feedback loop, which is discussed in detail in~\cite{VG-TR13}.
Thus, conditions use generic information about \EUREMA concepts to branch the \mbox{control}~flow while the different exit compartments of operations may depend on information only known internally to user-defined runtime models and operation implementations.
Model operations work on runtime \emph{models} represented by rectangles and the \emph{usage of models} as input or output is depicted by dotted arrows. Finally, an \emph{\FeedbackLoopDiagram model} can be used as a runtime model within another \FeedbackLoopDiagram, which will be discussed in \mbox{Section}~\ref{sec:ld:layered-architecture}.

To support the engineer's perception of \FeedbackLoopDiagrams, the elements can be substantiated by labels or stereotypes. 
Model operations are assigned to the typical steps of a feedback loop: 
\elem{$\ll$Monitor$\gg$},\,\elem{$\ll$Analyze$\gg$},\,\elem{$\ll$Plan$\gg$}, and 
\elem{$\ll$Execute$\gg$}. 
Models are stereotyped based on the purpose they serve in self-adaptive software, which resulted from the categorization of runtime models discussed in Section~\ref{sec:requirements}: 
\elem{$\ll$MonitoringModel$\gg$},
\elem{$\ll$ExecutionModel$\gg$},
\elem{$\ll$CausalConnectionModel$\gg$},
\elem{$\ll$ReflectionModel$\gg$},
\elem{$\ll$EvaluationModel$\gg$},
\elem{$\ll$ChangeModel$\gg$}, and
\elem{$\ll$AdaptationModel$\gg$}.
Finally, the usage of models by operations is substantiated to 
\elem{\textbf{c}}reating,
\elem{\textbf{d}}estroying,
\elem{\textbf{w}}riting,
\elem{\textbf{r}}eading, and
\elem{\textbf{a}}nnotating
models.
While reading a model does not have any side effects, writes modify the model in a way that potentially affects the adaptable software, and annotations to a model enrich a model without affecting the software.

With the \FeedbackLoopDiagram of Figure~\ref{fig:mm:self-repair}, we modeled the extended self-repair scenario from~\cite{VG10}. 
The \elem{Update} and \elem{Effect} operations use triple graph grammar rules (\elem{TGG Rules}) that specify by means of model transformation rules how the \elem{Architectural Model} reflecting the adaptable software (mRUBiS) is synchronized with the software. Thus, monitoring the software, the \elem{Update} operation keeps the \elem{Architectural Model} up-to-date. The \elem{Check for failures} operation performs analysis by applying \elem{Failure analysis rules} on the \elem{Architectural Model}. These rules define checks to identify critical failures. If no failures are found, the feedback loop terminates in the state \elem{Analyzed}. Otherwise, adaptation is required to repair them. 
At first, a decision is made whether further analysis is needed. This is the case when the condition holds, which checks whether the last execution of the \elem{Check for failures} operation that has identified \elem{no failures} happened more than five consecutive executions in the past. Thus, the past five runs of the loop did not repair the failures. The plan activity uses the analysis results annotated by the analyze operations to the \elem{Architectural Model} to select suitable \elem{Repair strategies}. The selected strategies change the \elem{Architectural Model} to prescribe a reconfiguration of the software. This reconfiguration is executed by the \elem{Effect} operation that synchronizes the \elem{Architectural Model} changes to the software, which terminates one run of the feedback~loop.

This example illustrates how adaptation activities can be considered as abstract model operations working on runtime models. Besides the control flow between the operations, the interplay between operations and runtime models is made explicit as the models are the basis for coordinating the activities. Thus, this interplay is similar to dependencies among adaptation activities that are relevant for properly specifying and executing a feedback loop.

\subsection{Modularizing Feedback Loop Diagrams}\label{sec:seams2012:modularFLD}

Besides modeling a feedback loop in a single \FeedbackLoopDiagram, \EUREMA supports modular specifications. Individual adaptation activities are specified in distinct \FeedbackLoopDiagrams that can be composed to form a feedback loop. This provides further abstractions that ease the modeling and perception of feedback loops by engineers. Parts of a loop can be abstracted in dedicated \FeedbackLoopDiagrams and referenced by other \FeedbackLoopDiagrams. This additionally supports reusability of such parts.

\begin{figure}[t]
\hspace{0.01\linewidth}
\begin{minipage}[b]{0.45\linewidth}
\centerline{\includegraphics[scale=.264]{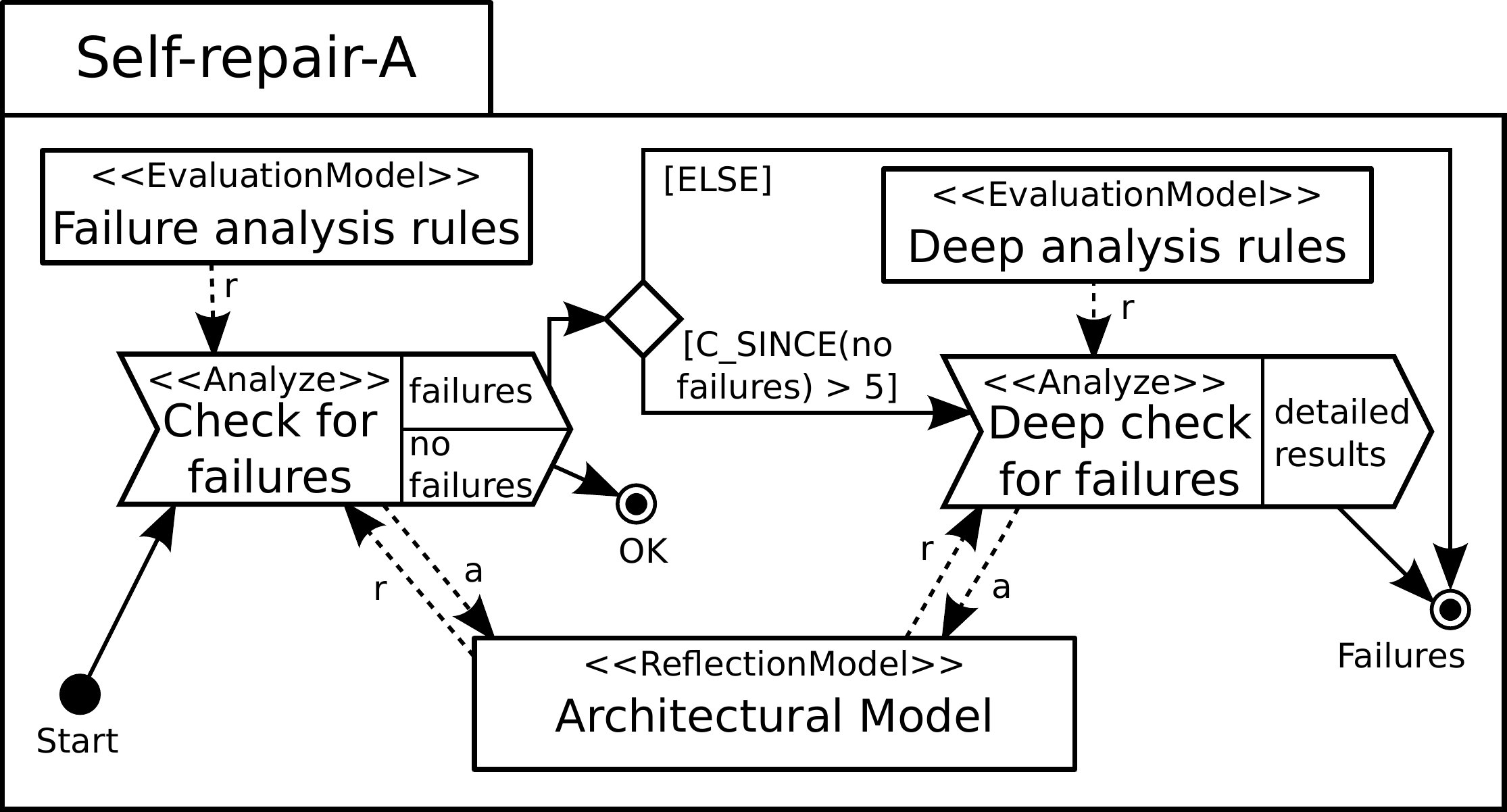}}
\caption{\FeedbackLoopDiagram for the self-repair analysis}
\label{fig:mm:self-repair-A}
\end{minipage}%
\hspace{0.01\linewidth}
\begin{minipage}[b]{0.49\linewidth}
\centerline{\includegraphics[scale=.264]{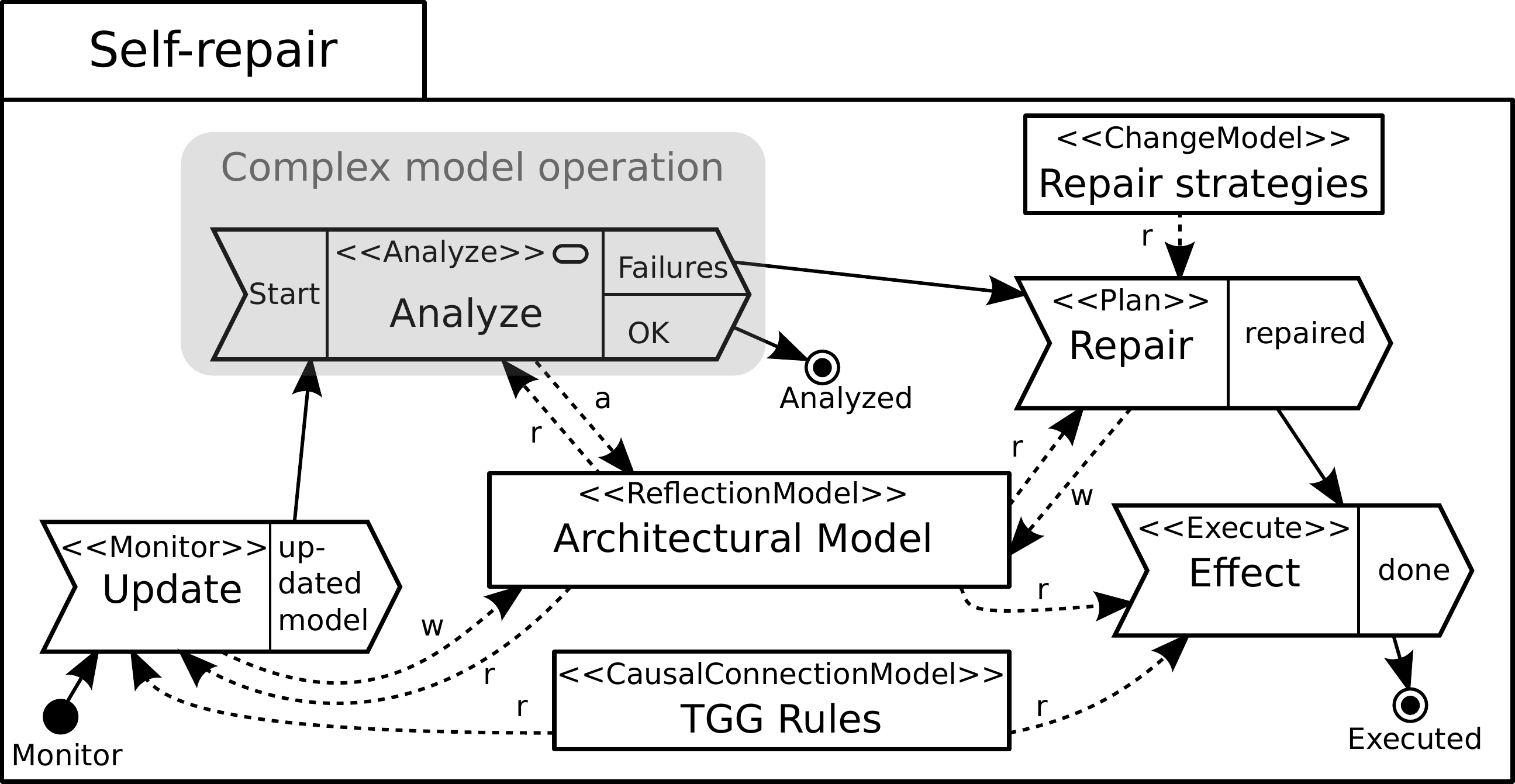}}
\caption{\FeedbackLoopDiagram invoking the {\sf Self-repair-A} \FeedbackLoopDiagram}
\label{fig:mm:self-repair-withoutA}
\end{minipage}
\hspace{0.01\linewidth}
\end{figure}

For instance, the analyze activity of the self-repair feedback loop (cf. Figure~\ref{fig:mm:self-repair}) can be specified in an \FeedbackLoopDiagram called \elem{Self-repair-A} (cf. Figure~\ref{fig:mm:self-repair-A}). This activity has one initial state (\elem{Start}) and two final states reflecting whether failures have been identified (\elem{Failures}) or not (\elem{OK}).
This \FeedbackLoopDiagram can be (re)used by other \FeedbackLoopDiagrams such as the \elem{Self-repair} loop shown in Figure~\ref{fig:mm:self-repair-withoutA}. Therefore, we introduce the concept of a \emph{complex model operation}. Such an operation defines a signature to synchronously invoke an \FeedbackLoopDiagram by referring to the initial and final states of the \FeedbackLoopDiagram. In the example, based on the initial and final states of the \elem{Self-repair-A} \FeedbackLoopDiagram, the complex model operation \elem{Analyze} in Figure~\ref{fig:mm:self-repair-withoutA} has one entry (\elem{Start}) and two exit compartments (\elem{Failures} and \elem{OK}). Thus, initial and final states of an \FeedbackLoopDiagram are mapped to entry and exit compartments, respectively. 
This ensures that the feedback loop using a complex operation can properly invoke an \FeedbackLoopDiagram and properly resume execution after the invocation. If an \FeedbackLoopDiagram specifies exactly one initial or final state, the entry or exit points for execution are uniquely defined and the entry or exit compartments of the complex operation can be omitted. In the example, the entry compartment of the complex model operation could have been omitted.

Moreover, the complex operation \elem{Analyze} uses the \elem{Architectural Model} (cf. Figure~\ref{fig:mm:self-repair-withoutA}) that can be considered as a parameter when invoking the \elem{Self-repair-A} \FeedbackLoopDiagram while the other runtime models used in the \elem{Self-repair-A} \FeedbackLoopDiagram are encapsulated by this \FeedbackLoopDiagram.
Considering self-adaptive software that evolves throughout its lifetime, \EUREMA adopts a dynamic typing approach in contrast to a static and explicit type system for \FeedbackLoopDiagrams and operations.
Finally, a complex operation is labeled with an icon to distinguish it from  basic model operations in \FeedbackLoopDiagrams and to reveal that it invokes another \FeedbackLoopDiagram. Thus, a complex operation used in an \FeedbackLoopDiagram abstracts from another \FeedbackLoopDiagram and it synchronously invokes the abstracted \FeedbackLoopDiagram when being executed.

Overall, the specifications of the self-repair feedback loop as the \FeedbackLoopDiagrams of Figure~\ref{fig:mm:self-repair} or Figures~\ref{fig:mm:self-repair-A} and \ref{fig:mm:self-repair-withoutA} are equivalent considering functionality. The only difference is the number of used \FeedbackLoopDiagrams, which is motivated by design decisions concerning abstraction and modularity. Besides the analyze activity, each of the four MAPE activities can be specified in distinct \FeedbackLoopDiagrams and a high-level \FeedbackLoopDiagram integrates them to a feedback loop. In general, the depth of abstraction and invocation relationships is not restricted. This leverages different abstraction levels for modeling and further assists engineers in understanding feedback~loops.

\subsection{Modeling Multiple Feedback Loops and their Coordination}\label{sec:seams2012:multipleFLD}

\begin{figure}
	\centerline{\includegraphics[scale=.264]{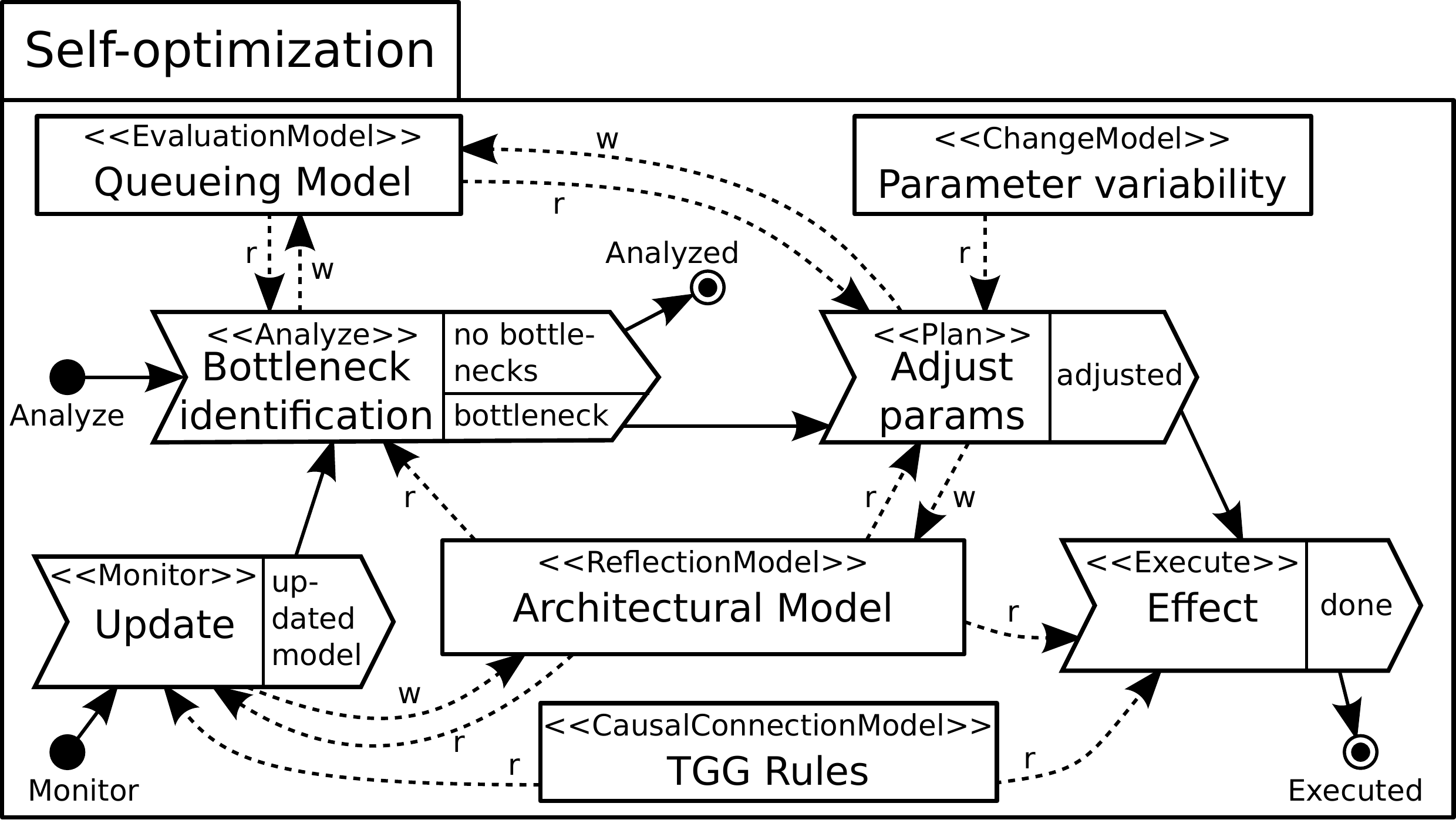}}
	\caption{\FeedbackLoopDiagram for {\sf Self-optimization}}
	\label{fig:mm:self-optimization}
\end{figure}

We discuss now how \EUREMA supports multiple feedback loops in self-adaptive software, particularly to handle multiple concerns such as self-repair or self-optimization. Each concern is managed by an individual feedback loop since each concern requires its specific runtime models and operations. 
Therefore, we extend our example of the self-adaptive software employing a self-repair feedback loop (cf. Section~\ref{sec:seams2012:modularFLD}) with a self-optimization feedback loop as specified by the \FeedbackLoopDiagram in Figure~\ref{fig:mm:self-optimization}. Similar to the self-repair loop, the \elem{Update} and \elem{Effect} operations synchronize the \elem{Architectural Model} with the adaptable software. Besides the \elem{Architectural Model}, the analyze and plan activities~use~a \elem{Queueing Model} to identify bottlenecks in the adaptable software and reasonable values for parameters given by the \elem{Parameter variability} model to adjust the configuration~of~the software, which aims for resolving the bottlenecks. Finally, the self-optimization loop has two initial states either initiating the loop with the monitor or the analyze activity.

In general, \EUREMA supports the specification of multiple feedback loops by distinct \FeedbackLoopDiagrams. However, employing multiple loops raises questions of possible interferences. Concerns such as failures and performance are typical competing, like healing a failure results in a system with decreased performance. Such interferences require coordination.
In \EUREMA, the coordination of multiple feedback loops is explicitly modeled with an \FeedbackLoopDiagram that synchronizes the execution of the feedback loops. In the following, we discuss two basic design alternatives for coordinating the self-repair and self-optimization feedback loops.

\subsubsection{Sequencing Complete Feedback Loops}\label{sec:seams2012:multipleFLD:self-management-1}

\begin{figure}[t]
\centerline{\includegraphics[scale=.264]{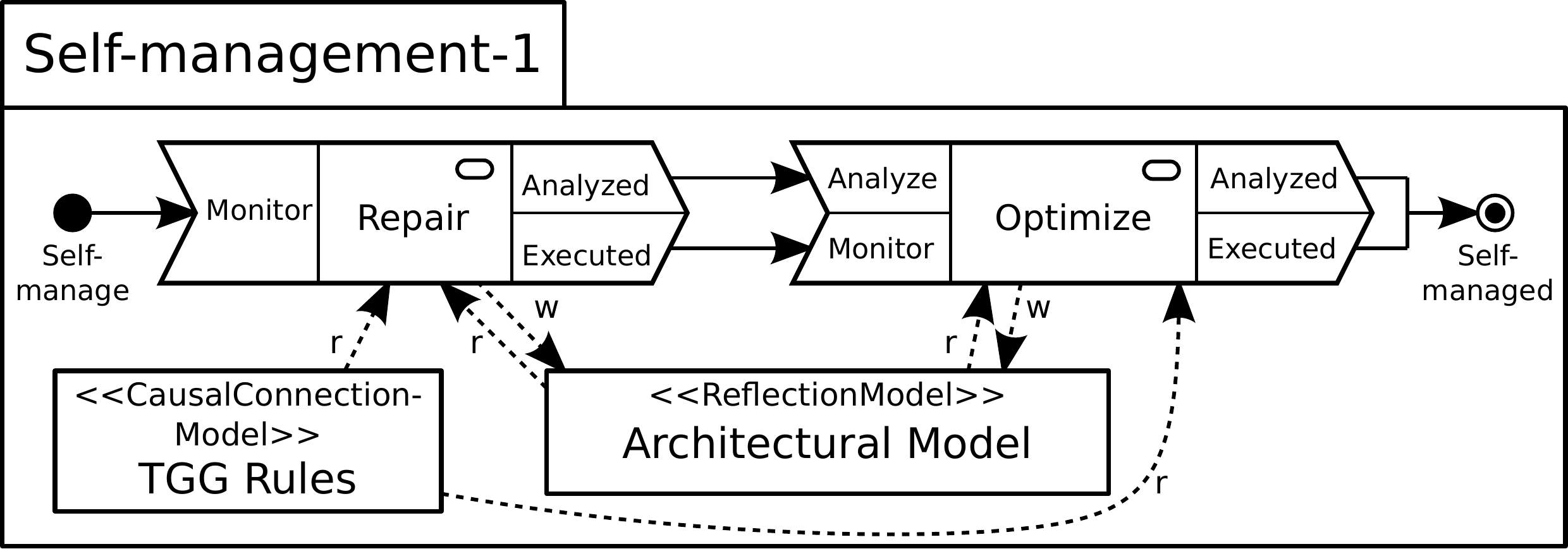}}
\caption{\FeedbackLoopDiagram for {\sf Self-management-1}}
\label{fig:mm:loop-sequence}
\vspace{1em}
\end{figure}

A simple way to coordinate two feedback loops is to execute them sequentially. This is specified by the \elem{Self-management-1} \FeedbackLoopDiagram in Figure~\ref{fig:mm:loop-sequence} that uses complex model operations to synchronously invoke the individual feedback loops. In this self-management example, a higher priority is assigned to repairing failures than to optimizing performance since optimizing the performance of a failing system is not reasonable. Therefore, the self-repair feedback loop is executed before the self-optimization~loop.

In the \FeedbackLoopDiagram, \elem{Repair} invokes the self-repair loop (Figure~\ref{fig:mm:self-repair-withoutA}) to start execution in the state \elem{Monitor}. Thus, the monitor and analyze activities are carried out to update and check the \elem{Architectural Model} for failures. If no failures are identified, the self-repair loop~does not plan and execute any adaptation and terminates in the state \elem{Analyzed}. In~this case, the subsequent self-optimization loop may immediately start with the \mbox{analyze} activity because the self-repair's monitor has already updated the shared \elem{Architectural Model}. Thus, the complex operation \elem{Optimize} invokes the self-optimization loop (Figure~\ref{fig:mm:self-optimization}) to begin execution in the state \elem{Analyze} (cf. the control flow connecting the \elem{Repair} and \elem{Optimize} operations). If no bottlenecks have been identified, the self-optimization feedback loop terminates in the state \elem{Analyzed}. Otherwise, it performs the plan and execute activities, and terminates in the state \elem{Executed}.

On the other hand, if the self-repair feedback loop identifies failures, it plans and executes an adaptation to the adaptable software and terminates in the state \elem{Executed}. This requires that the subsequent self-optimization feedback loop (Figure~\ref{fig:mm:self-optimization}) performs monitoring to observe the effects of this adaptation. Therefore, \elem{Optimize} invokes the self-optimization loop to begin execution in the state \elem{Monitor} (cf. the control flow connecting the \elem{Repair} and \elem{Optimize} operations). After carrying out the monitor and analyze activities, the self-optimization feedback loop either terminates or, if required, plans and executes an adaptation.

This coordination design synchronizes different feedback loops by sequentially executing them based on priorities and by using the adaptable software for synchro\-nization. Thus, an adaptation performed by one feedback loop is executed to the \mbox{software} before another loop starts execution with the monitor activity to observe the adaptation effects. However, if a feedback loop does not perform any adaptation of the~reflection model and software, the subsequent loop may skip the monitoring and start with the analyze activity since the previous loop already performed the monitoring to update the shared reflection model.

\subsubsection{Sequencing Adaptation Activities of Feedback Loops}\label{sec:seams2012:multipleFLD:self-management-2}

\begin{figure}[t]
	\centerline{\includegraphics[scale=.264]{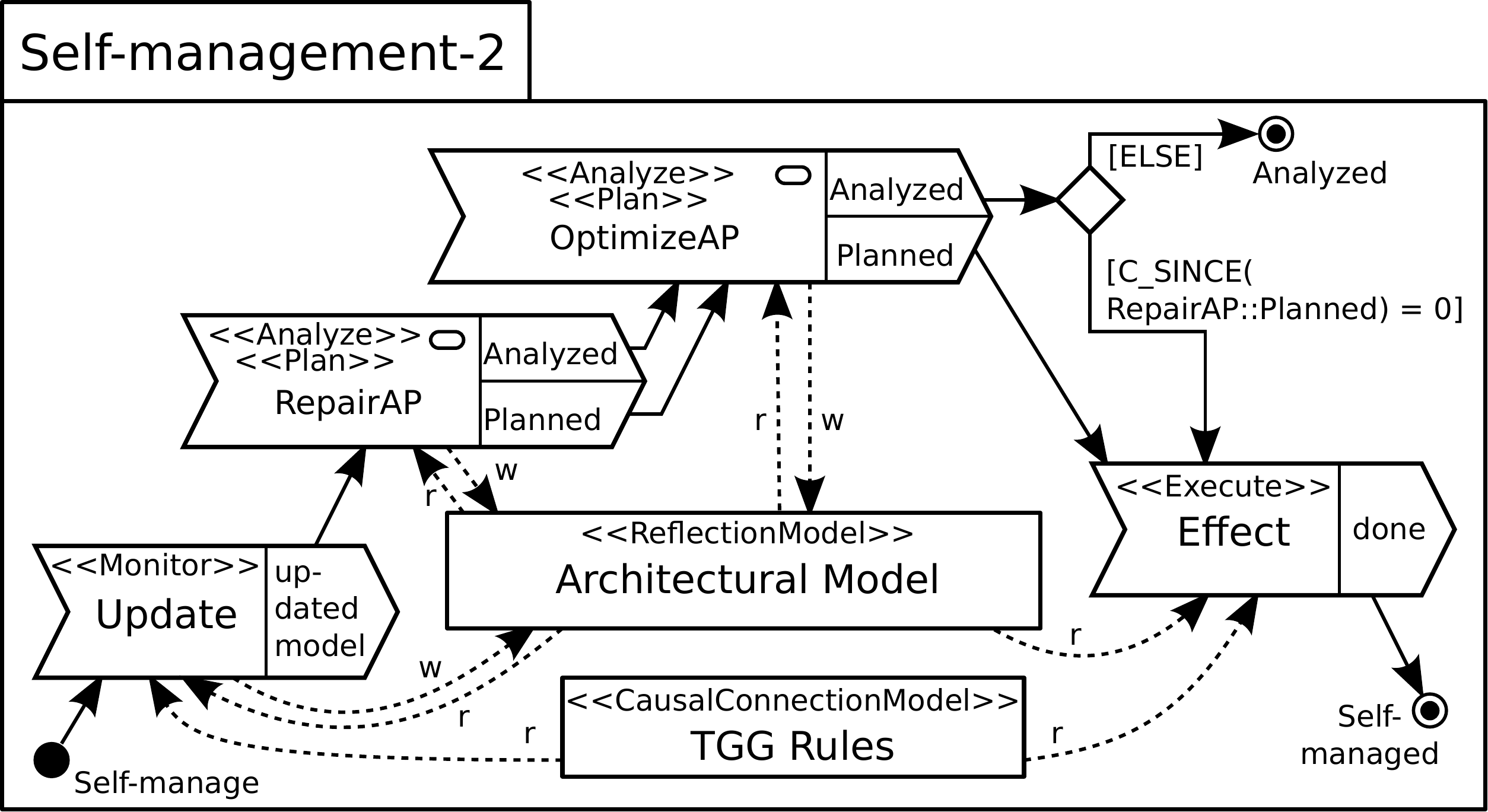}}
	\caption{\FeedbackLoopDiagram for {\sf Self-management-2}}
	\label{fig:mm:loop-overlapping}
\vspace{-.5em}
\end{figure}

\begin{figure}[t]
	\centering
	\subfloat{\includegraphics[scale=.26]{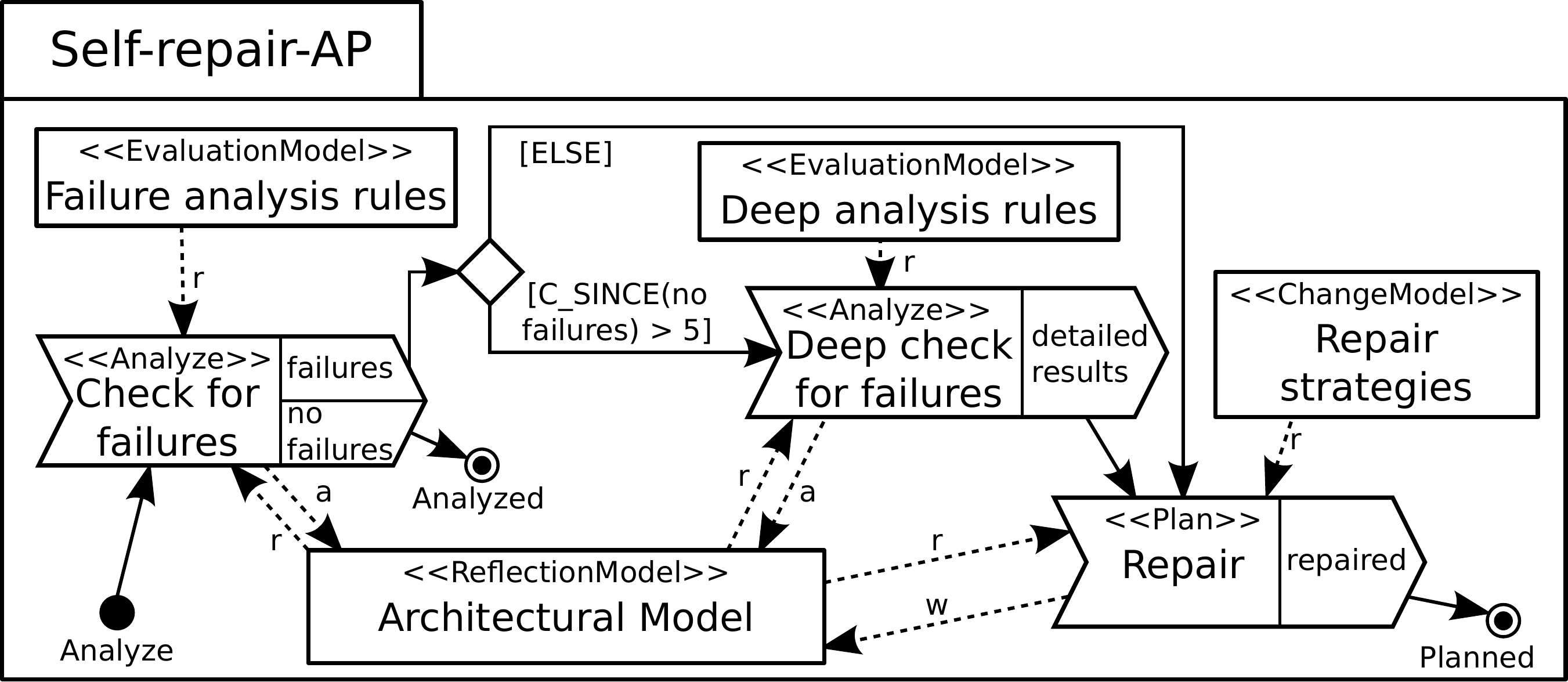}}
	\hspace{0.02\columnwidth}
	\subfloat{\includegraphics[scale=.26]{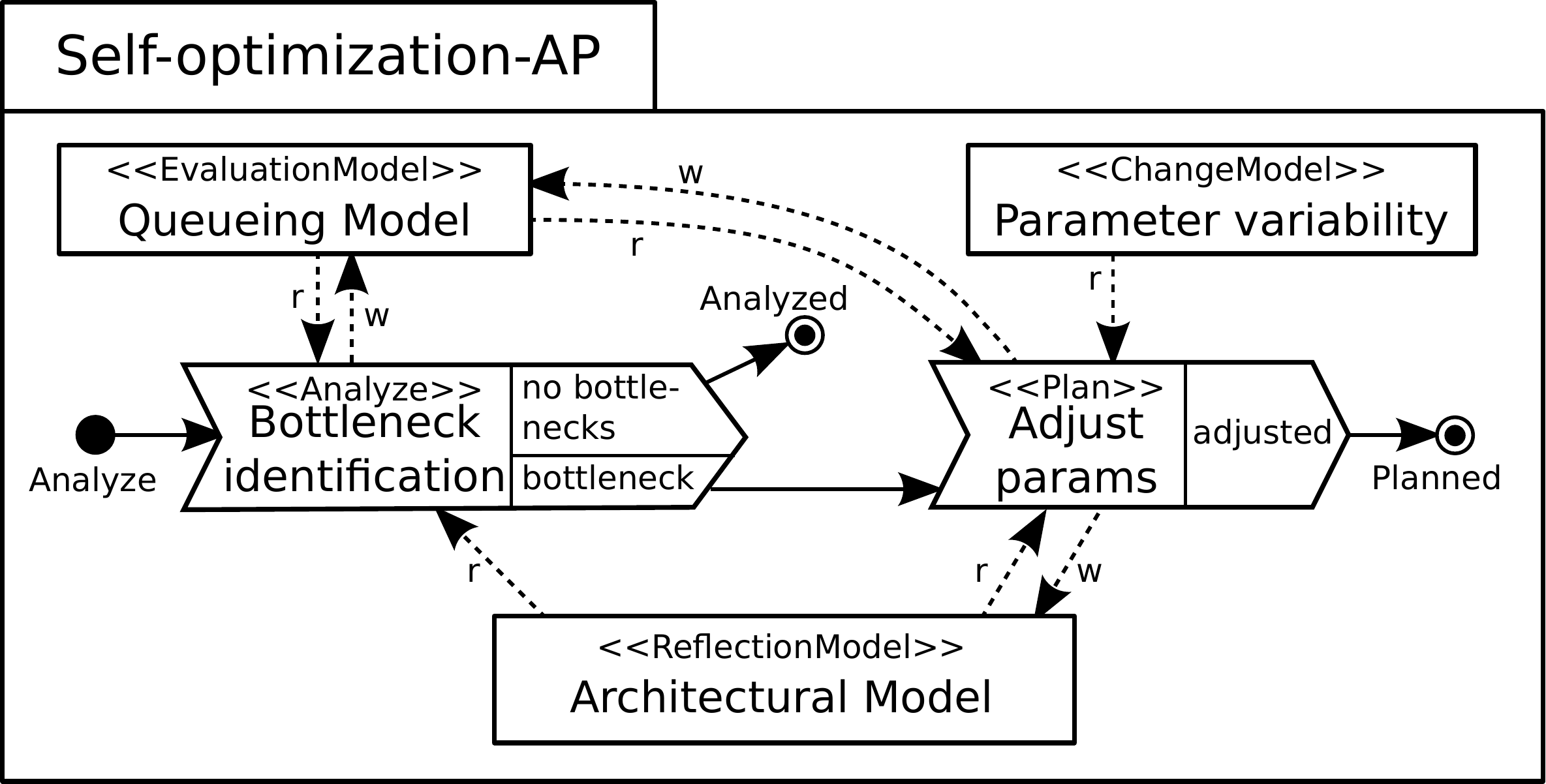}}
	\caption{Analyze and plan activities of the self-repair (left) and the self-optimization (right)}
	\label{fig:mm:AP}
\end{figure}

The other design alternative for coordinating multiple feedback loops synchronizes them in shared monitor and execute activities and sequentially executes the individual analyze and plan activities. 
For the example, this is specified by the \mbox{\elem{Self-management-2}} \FeedbackLoopDiagram in Figure~\ref{fig:mm:loop-overlapping}. The \elem{Update} and \elem{Effect} operations synchronize the \elem{Architectural Model} with the adaptable software. This model is shared by the individual, concern-specific analyze and plan activities of the self-repair and self-optimization loops (cf. complex operations \elem{RepairAP} and \elem{OptimizeAP}) that are specified in individual \FeedbackLoopDiagrams shown in Figure~\ref{fig:mm:AP}. 
Thus, analysis and planning for the self-repair are executed before the analysis and planning for the self-optimization. The \elem{Architectural Model} is only modified by the self-repair's plan activity if the related analyze activity has identified failures. These modifications are a planned adaptation to repair the failures, i.e., they are applied in the model but not effected to the adaptable software. If the model has not been modified, there are no conflicting adaptations possible. 
Otherwise, the adaptations proposed for the self-repair must be handled by the subsequent analyze and plan activities for the self-optimization, which requires coordination.

While \EUREMA coordinates the \emph{execution} of the individual activities/operations, in this case by strictly sequencing them for self-repair and self-optimization, the implementations of these operations are responsible for establishing a \emph{consensus} on conflicting adaptations, in this case by considering the adaptation proposed for self-repair as an invariant for the self-optimization. In general, a consensus can be reached, for example, by mechanisms based on utility functions \cite{Cheng2006} or concepts of coordination and agreement in distributed systems such as voting \cite{Coulouris+2011}.
Such mechanisms are handled by the implementations of the model operations as they typically depend on internals of user-defined runtime models that are black boxes for \EUREMA. Thus, the concrete mechanisms are transparent for \EUREMA that focuses on providing execution support for these operations.

Considering the \FeedbackLoopDiagram in Figure~\ref{fig:mm:loop-overlapping}, when the self-optimization's analyze and plan activities terminate, the \elem{Effect} operation is executed if adaptations are proposed in the \elem{Architectural Model} by the self-repair's or the self-optimization's plan activities. Thus, at least one of the complex model operations \elem{RepairAP} or \elem{OptimizationAP} must terminate in the state \elem{Planned}. Otherwise, the \elem{Self-management-2} module terminates in the state \elem{Analyzed} because no critical failures and no bottlenecks have been identified, which does not require any adaptation.

\section{Layer Diagrams}\label{sec:ld}

While \FeedbackLoopDiagrams support the behavioral specification and coordinated execution of feedback loops, \EUREMA so far does not structurally reflect all \FeedbackLoopDiagrams employed in a self-adaptive software as well as their interrelationships and triggering. Therefore, we extend \EUREMA in this article with \emph{layer diagrams} (\LayerDiagrams) that provide an architectural view. This view refers to an instance situation of the software and it captures the employed feedback loops, their relationships to each other and to the adaptable software, and their triggering conditions.

\begin{figure}[t]
	\hspace{0.01\linewidth}
	\begin{minipage}[b]{0.45\linewidth}
		\centerline{\includegraphics[scale=.3]{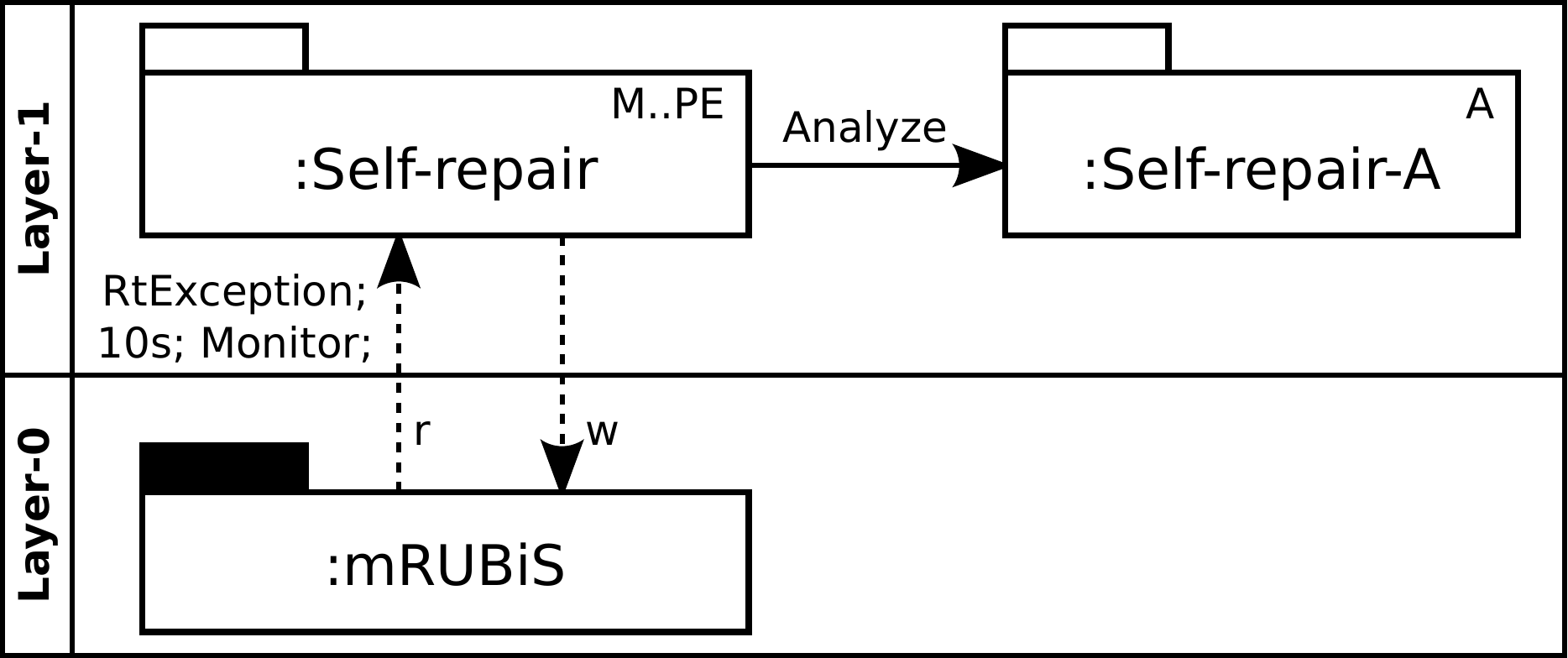}}
		\caption{\LayerDiagram for {\sf Self-repair}}
		\label{fig:mm:self-repair-layer}
	\end{minipage}%
	\hspace{0.04\linewidth}
	\begin{minipage}[b]{0.45\linewidth}
		\centerline{\includegraphics[scale=.314]{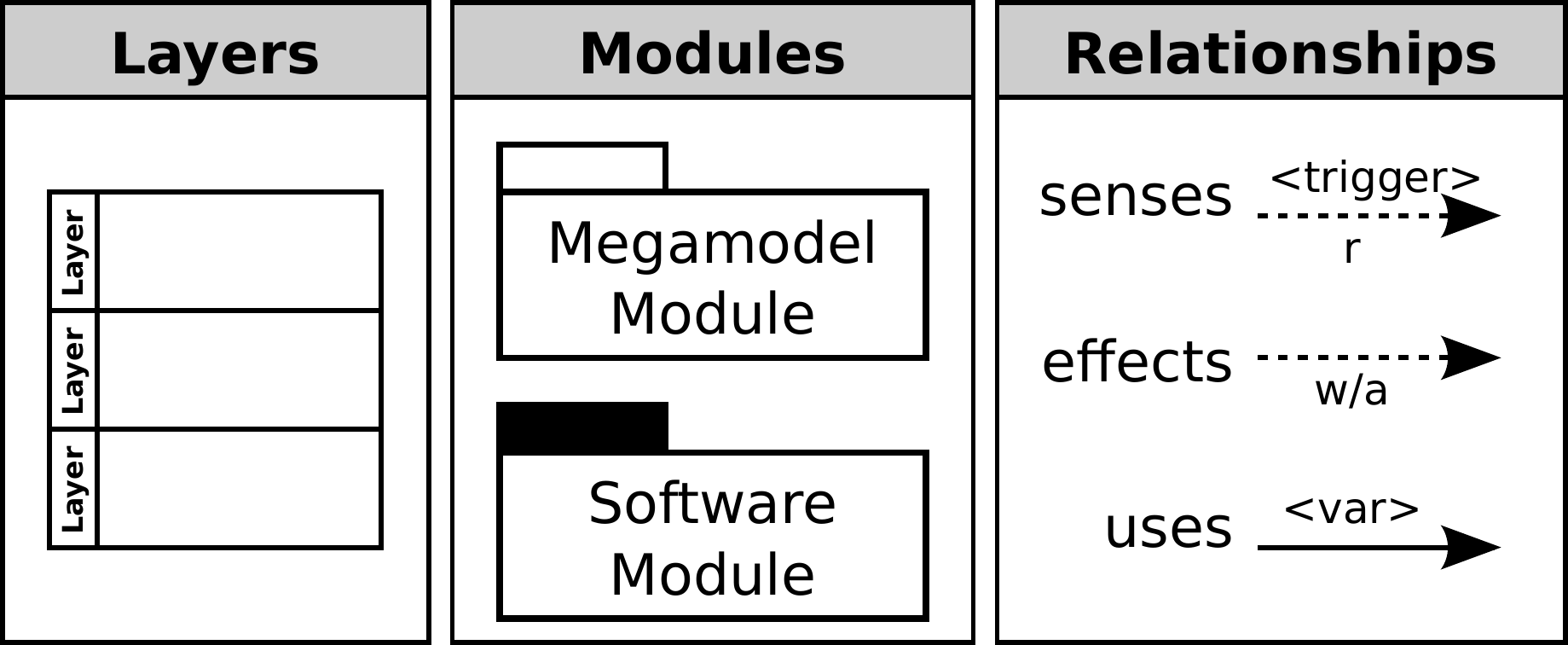}}
		\caption{Concrete syntax for \LayerDiagrams}
		\label{fig:concrete-syntax-LD}
	\end{minipage}
	\hspace{0.01\linewidth}
\end{figure}

This is exemplified for the self-repair example by the \LayerDiagram in Figure~\ref{fig:mm:self-repair-layer}. It specifies that an instance of the \elem{Self-repair} feedback loop as specified by the corresponding \FeedbackLoopDiagram (cf. Figure~\ref{fig:mm:self-repair-withoutA}) is located at \elem{\mbox{Layer-1}} and directly senses and effects the running \elem{:mRUBiS} instance at \elem{\mbox{Layer-0}}. 
As defined by the concrete syntax (cf. Figure~\ref{fig:concrete-syntax-LD}), partitions in \LayerDiagrams represent the \emph{layers} of the self-adaptive software, which contain \emph{modules}. An instance of an \FeedbackLoopDiagram constitutes a \emph{megamodel module} that encapsulates the details of a concrete feedback loop or adaptation activities. Such a module is depicted as a package with a white tab in an \LayerDiagram since \EUREMA provides white-box views of megamodel modules by means of the \FeedbackLoopDiagrams. In contrast, a package with a black tab represents \emph{software modules} that are not specified by \EUREMA, like an instance of the adaptable software. \emph{Sense} and \emph{effect relationships} between modules are reflected by dotted arrows, either labeled with \elem{r} (for \textbf{r}eading the sensed module) or \elem{w}/\elem{a} (for \textbf{w}riting/\textbf{a}nnotating the effected module), respectively. A sense relationship is labeled with a triggering condition for the sensing module, which will be discussed in Section~\ref{sec:ld:trigger}.

Usage relationships between megamodel modules are modeled in \LayerDiagrams to make the dependencies between \FeedbackLoopDiagrams explicit. The example \LayerDiagram (cf. Figure~\ref{fig:mm:self-repair-layer}) explicitly shows that the \elem{:Self-repair} module defined by the \FeedbackLoopDiagram in Figure~\ref{fig:mm:self-repair-withoutA} uses the \elem{\mbox{:Self-repair-A}} module defined by the \FeedbackLoopDiagram in Figure~\ref{fig:mm:self-repair-A}. A \emph{use} relationship is reflected by a solid arrow and it is labeled with the name of the corresponding complex model operation (\elem{Analyze} in this example) to bind this operation of the invoking module (\elem{:Self-repair} in this example) to the concrete module (\elem{:Self-repair-A} in this example) to be invoked. 
The same mechanism is used to bind basic model operations to their implementations that are black boxes for \EUREMA and therefore modeled as \emph{software modules} in \LayerDiagrams. An \FeedbackLoopDiagram specifies when a basic model operation should be executed, which runtime models are used as input and output, and the operation's return states while the concrete implementations of such operations have to be provided by software modules. When executing a basic operation as part of a megamodel module (\FeedbackLoopDiagram instance), \EUREMA invokes the software module to which the operation is bound by a \emph{use relationship} in the \LayerDiagram and labeled with the operation name.
So far and in the following examples, we omit the modeling of software modules implementing basic model operations in \LayerDiagrams since they are just required for binding the operations to their implementations. This is relevant for the execution but they provide no further information relevant for the design of feedback loops.

Finally, the \LayerDiagram shows by the labels \elem{M..PE} and \elem{A} that the \elem{:Self-repair} module realizes the monitor, plan, and execute activities and the \mbox{\elem{:Self-repair-A}} module the analyze activity. Inspired by \citeN{SEfSAS2-decentral}, we use such labels to indicate which adaptation activities are realized by a megamodel module as defined in an \FeedbackLoopDiagram.
Thus, an \FeedbackLoopDiagram instance encapsulates the behavioral specification of a feedback loop or individual adaptation activities in a megamodel module. The \LayerDiagram considers such modules as black boxes and describes in which layers they are located and the relationships to other megamodel or software modules. Therefore, an \LayerDiagram provides an abstract and structural view, which supports the architectural design of self-adaptive software and complements the \FeedbackLoopDiagrams as behavioral views of feedback loops.

\subsection{Triggering Conditions for Feedback Loops}\label{sec:ld:trigger}

Besides specifying a feedback loop with \FeedbackLoopDiagrams, a triggering condition is required to determine \emph{when} an instance of the feedback loop should be executed. In \EUREMA, we especially consider the occurrences of events as triggers. These events are emitted by modules that are either adaptable software or feedback loop instances. A triggering condition for a feedback loop instance may only refer to events emitted from those modules that are sensed by this instance. This avoids that the feedback loop instance is triggered by events related to modules that are of no interest to the instance. Thus, we specify triggering conditions in \LayerDiagrams by annotating them to the corresponding \emph{sense} relationships that reveal the flow of events from one module to another module.
This is exemplified by the \LayerDiagram in Figure~\ref{fig:mm:self-repair-layer} that defines a triggering condition for the \elem{:Self-repair} feedback loop sensing the \elem{:mRUBiS} system. The condition \elem{RtException; 10s; Monitor;} defines that the \elem{:Self-repair} loop starts execution if the \elem{:mRUBiS} system emits an event of type \elem{RtException} notifying about a runtime exception, and when ten seconds since the last execution of this loop have expired. Finally, \elem{Monitor} points to the initial state of the feedback loop (cf.~Figure~\ref{fig:mm:self-repair-withoutA}), in which the execution starts. 

\EUREMA supports a simple language for triggering conditions consisting of three parts: \elem{events;}\,\elem{period;} \elem{initialState;}. 
The first part (\elem{events}) refers to a list of events that have to be modeled in \EUREMA. Therefore, the \EUREMA metamodel (cf. Section~\ref{sec:implementation}) provides the concepts of \elem{Event} and \elem{EventType} to model events in a hierarchy of event types. At runtime, these events as part of triggering conditions are matched against events actually emitted by modules. For instance, in our prototype we employ an infrastructure that makes software realized with \emph{Enterprise Java Beans 3} technology, such as our \mRUBiS system \cite{mRUBiS}, observable and adaptable and that emits events using the \emph{Java Messaging Service} (JMS) to notify about changes in the running software~\cite{VG10}. Such JMS events are then matched against the events modeled in \EUREMA and used in triggering conditions. If a match has been identified, the trigger of the feedback loop instance is activated.

The second part (\elem{period}) defines the minimal time period between two consecutive runs of the feedback loop instance, which is measured as the time elapsed between the termination of the previous run and the beginning of the next run. Thus, if the required event that activates the trigger occurs before the specified time period has elapsed, the next run will be delayed until the period eventually has elapsed. Delaying the execution avoids thrashing effects due to the proliferation of events and it allows the adaptation being executed by the previous run to take effect in the adaptable software. Likewise, selecting specific events in the first part of the triggering condition also serves as a filter that avoids the execution of a loop instance for every event emitted by the sensed module. 
Finally, the third part (\elem{initialState}) just refers to the initial state, in which the instance should start its execution.

The first two parts of a condition are optional, but one of them must be specified. If no \elem{events} are specified, the \elem{period} must be defined, which results in a trigger that periodically executes the feedback loop instance. If no \elem{period} is defined, the \elem{events} must be specified and the trigger executes the instance when the corresponding events have occurred and the current run of the instance has terminated.
In \EUREMA, a feedback loop instance is not reentrant, and therefore, any events that occur while the instance is running are queued. Thus, there are no concurrent executions of the same feedback loop instance.
While an event-driven trigger supports \emph{reactive} adaptation, a periodical trigger makes \emph{proactive} adaptation possible by executing the loop instance before the adaptable software emits any event.

\subsection{Variability Modeling}\label{sec:ld:variability}

\begin{figure}[t]
\hspace{0.001\linewidth}
\begin{minipage}[b]{0.4\linewidth}
\centerline{\includegraphics[scale=.293]{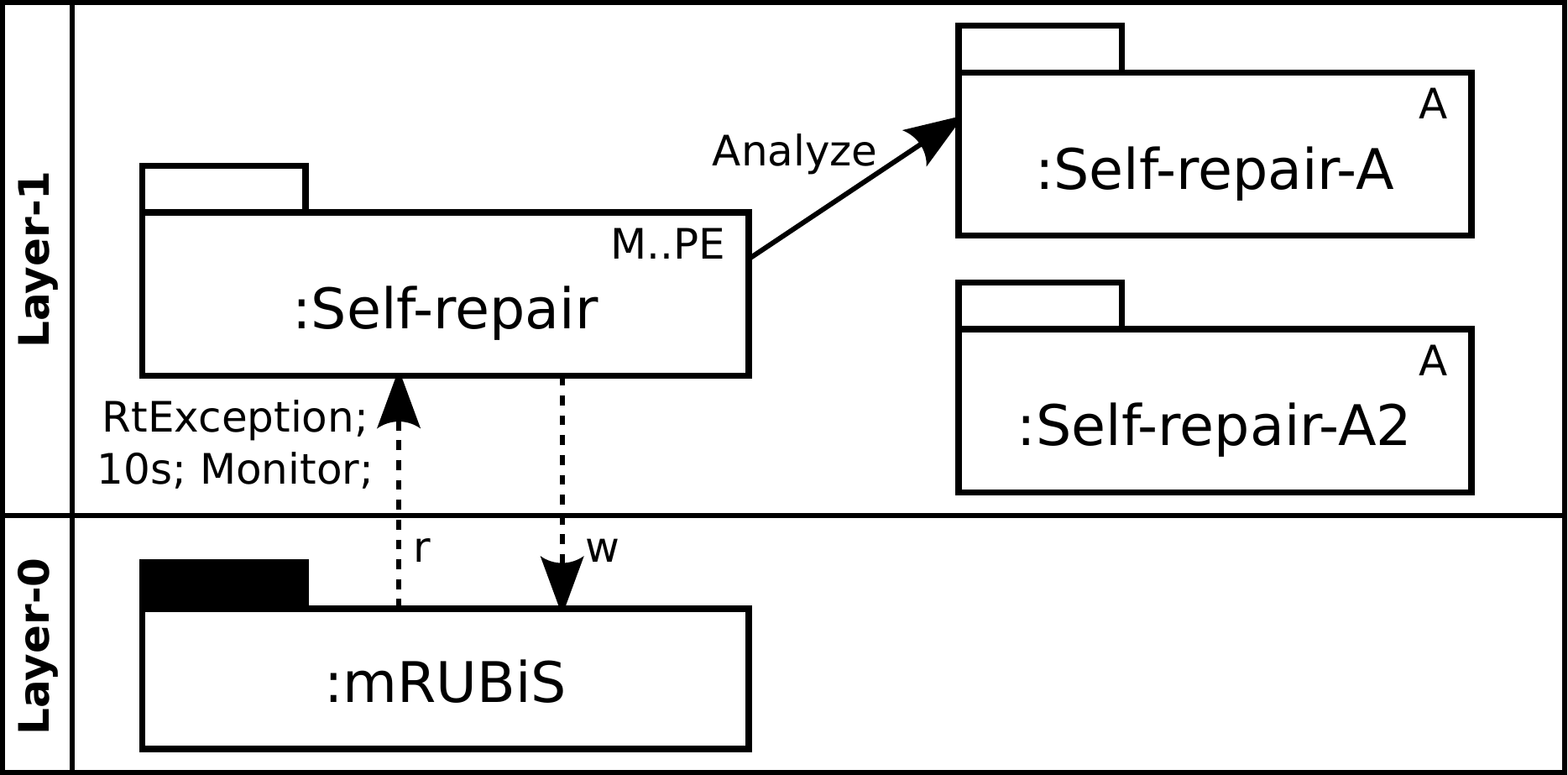}}
\caption{\LayerDiagram with variability}
\label{fig:mm:self-repair-layer-b-variation}
\end{minipage}%
\hspace{0.03\linewidth}
\begin{minipage}[b]{0.53\linewidth}
\centerline{\includegraphics[scale=.293]{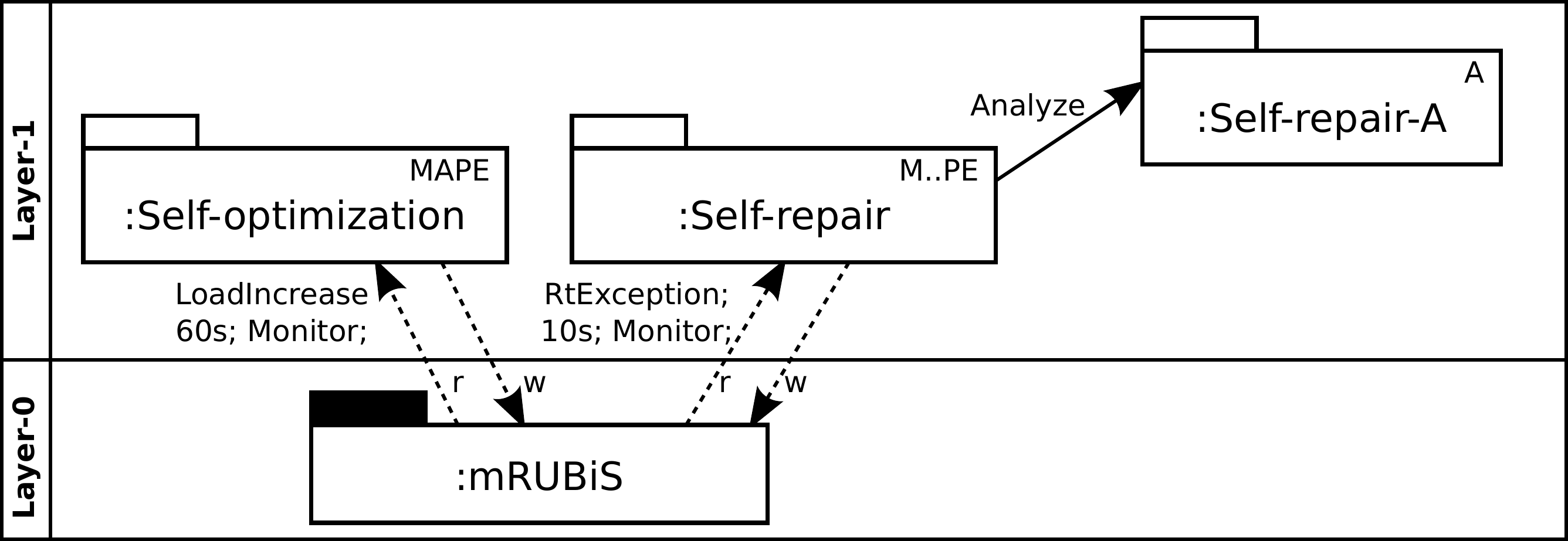}}
\caption{\LayerDiagram for the independent execution}
\label{fig:mm:multiple-loop-layer}
\end{minipage}
\end{figure}

Encapsulating feedback loops or individual adaptation activities defined by \FeedbackLoopDiagrams in megamodel modules and explicitly modeling such modules and their relationships in \LayerDiagrams reveals variation points in the adaptation engine. Such variability can be made visible in \LayerDiagrams and exploited to switch between variants, either during the design or dynamically at runtime.

For instance, having encapsulated the analyze activity of the self-repair feedback loop in a dedicated module (cf. \elem{:Self-repair-A} in the \LayerDiagram of Figure~\ref{fig:mm:self-repair-layer}), we assume that we have modeled an additional analyze activity (\elem{\mbox{Self-repair-A2}}) in a distinct \FeedbackLoopDiagram, which employs a different analysis technique than \elem{Self-repair-A}. Both activities are then alternative analyses to be used by the self-repair feedback loop. This constitutes a variation point in the architectural design reflected in the \LayerDiagram in Figure~\ref{fig:mm:self-repair-layer-b-variation}. If the \FeedbackLoopDiagrams of both alternatives have the same signature (cf. Section~\ref{sec:seams2012:modularFLD}), both of them can be used by the same complex model operation. Then, to switch between these alternatives, it is sufficient to change the binding between the complex model operation and the megamodel module, for example, by re-routing the \emph{use} relationship \elem{Analyze} to point to \elem{:Self-repair-A2} instead of \elem{:Self-repair-A} in the \LayerDiagram.

\EUREMA applies the same idea to leverage variability for basic model operations. \LayerDiagrams capture alternative software modules implementing a basic operation and the binding of the operation to one alternative. Similar to changing a binding of a complex operation to a megamodel module, the binding of a basic operation to a software module can be changed.

In general, such variations points of modules reify architectural variability of feedback loops in \LayerDiagrams. This variability can be exploited to specify and evaluate alternative feedback loop designs or to adjust feedback loops at runtime by switching between variants.

\subsection{Modeling Multiple Feedback Loops}\label{sec:ld:multiple-feedback-loops}

Employing multiple feedback loops in an adaptation engine, the relationships between these feedback loops should be captured, particularly concerning the execution. Therefore, it should be visible at the architectural level if multiple feedback loop instances are executed independently from each other or in a coordinated manner. Modeling the coordinated execution with \FeedbackLoopDiagrams has been discussed in Section~\ref{sec:seams2012:multipleFLD}, however, without considering the architectural level and the independent execution that cannot be explicitly covered by \FeedbackLoopDiagrams.

\subsubsection{Independent Execution of Multiple Feedback Loops}\label{sec:ld:multiple-feedback-loops:independent}

Based on the example of employing a self-repair and a self-optimization feedback loop (cf. Section~\ref{sec:seams2012:multipleFLD}), we assume that there are no interferences between these loops such that they can be executed independently. Thus, instances of both loops have individual triggers that might be activated at the same time and the instances might even run concurrently. This is specified by the \LayerDiagram in Figure~\ref{fig:mm:multiple-loop-layer} showing two independent feedback loop instances sensing and effecting the mRUBiS system.
While the triggering condition of the \elem{:Self-repair} module has been discussed in Section~\ref{sec:ld:trigger}, the \elem{:Self-optimization} module should be triggered in its initial state \elem{Monitor} when the load on the mRUBiS system increases, causing a \elem{LoadIncrease} event, and with a period of 60s. Based on the different triggering conditions, both modules run independently from each other and without any direct interactions as these modules are not interrelated  in the \LayerDiagram.

In general, an \LayerDiagram explicitly shows if multiple feedback loop instances are employed and executed independently. This would not be visible solely using \FeedbackLoopDiagrams as they do not reflect the triggering conditions and all of the modules employed in the adaptation engine.

\subsubsection{Coordinated Execution of Multiple Feedback Loops}\label{sec:ld:multiple-feedback-loops:coordinated}

Though the independent exe\-cution of multiple feedback loops is conceivable, there are sometimes interferences between them that have to handled by coordination. This has been discussed and modeled by \FeedbackLoopDiagrams in Section~\ref{sec:seams2012:multipleFLD}, which, however, do not make the employed feedback loop instances and their coordinated execution visible at the architectural level. We have discussed two basic mechanisms to coordinate two feedback loops, either by completely sequencing them (cf. Section~\ref{sec:seams2012:multipleFLD:self-management-1}) or by sequencing the individual analyze and plan activities (cf. Section~\ref{sec:seams2012:multipleFLD:self-management-2}). Both mechanisms are specified by \FeedbackLoopDiagrams (\elem{Self-management-1} and \elem{Self-management-2}, respectively) that invoke the individual \FeedbackLoopDiagrams defining the feedback loops by complex model operations. 

\begin{figure}[t]
\hspace{0.01\linewidth}
\begin{minipage}[b]{0.56\linewidth}
\centerline{\includegraphics[scale=.293]{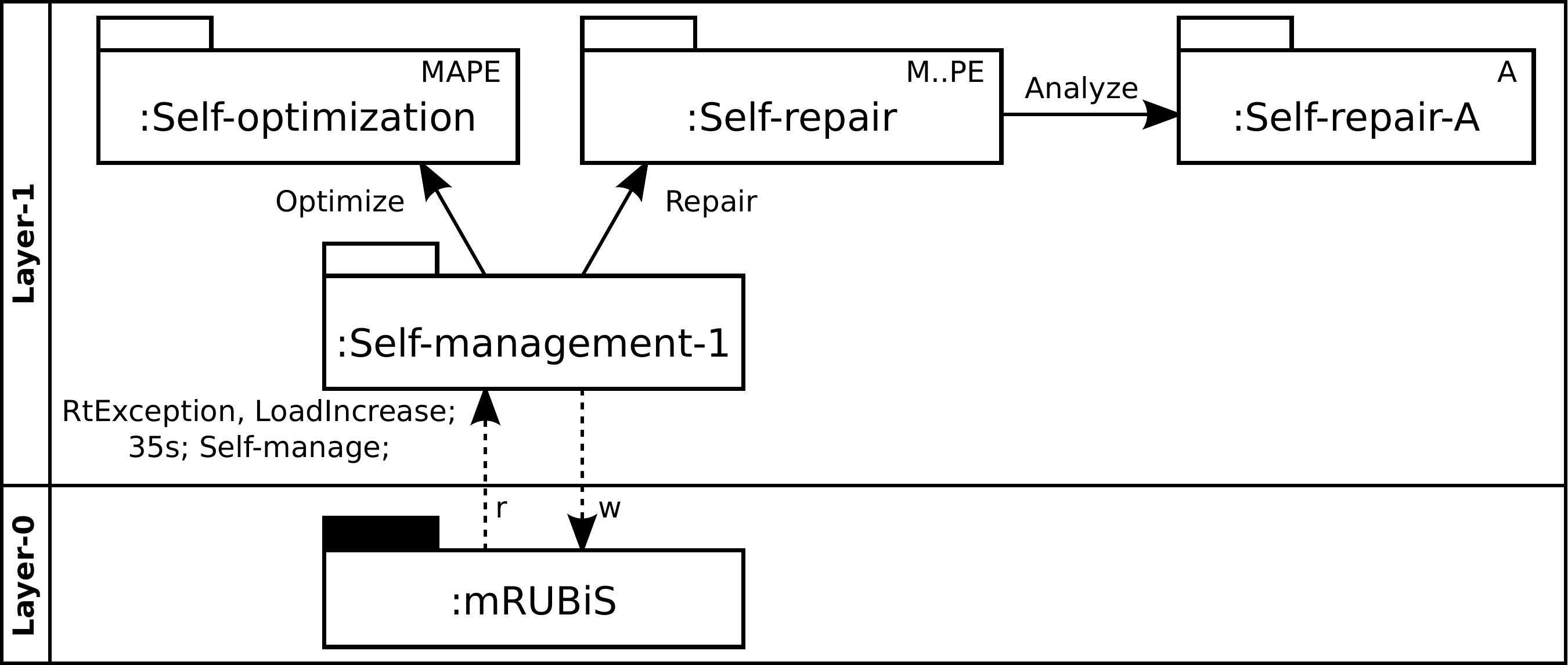}}
\caption{\LayerDiagram for {\sf Self-management-1}}
\label{fig:mm:loop-sequence-layer}
\end{minipage}%
\hspace{0.04\linewidth}
\begin{minipage}[b]{0.34\linewidth}
\centerline{\includegraphics[scale=.293]{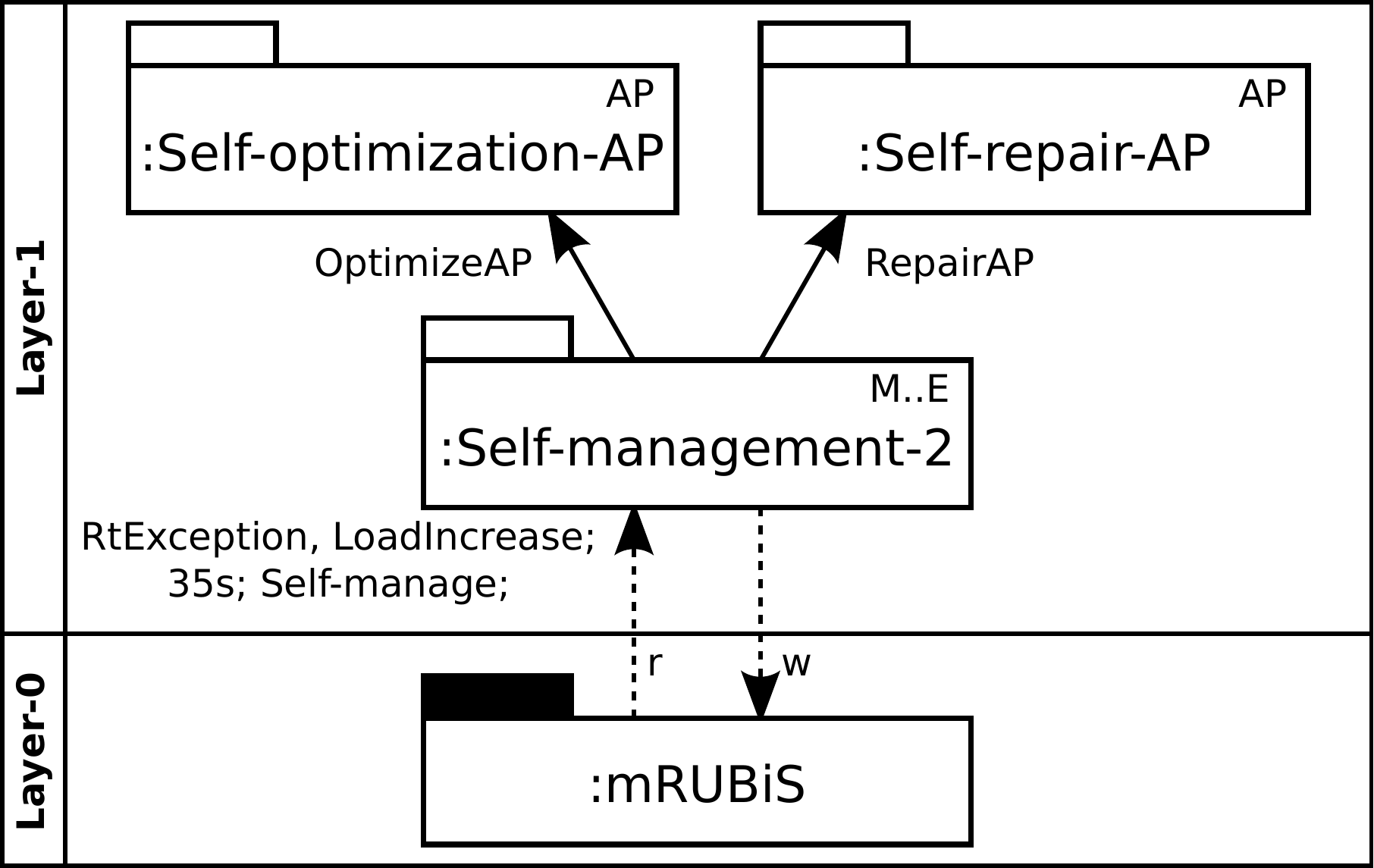}}
\caption{\LayerDiagram for {\sf Self-management-2}}
\label{fig:mm:loop-overlapping-layer}
\end{minipage}
\hspace{0.01\linewidth}
\end{figure}

To make this coordination explicit at the architectural level, \LayerDiagrams as shown in \mbox{Figures}~\ref{fig:mm:loop-sequence-layer} and~\ref{fig:mm:loop-overlapping-layer} are used. They show the modules whose execution is coordinated through invocations by the \elem{:Self-management-1} and \elem{:Self-management-2} modules. Therefore, the use relationships in the \LayerDiagrams make the invocation relationships among the modules explicit such as the \elem{Repair} relationship showing that the \elem{:Self-management-1} module invokes the \elem{:Self-repair} module. Moreover, the modules \mbox{\elem{:Self-management-1}} and \mbox{\elem{:Self-management-2}} realizing the coordinated execution require triggering conditions as shown in the \LayerDiagrams. These conditions combine the individual conditions of the self-repair and self-optimization feedback loops discussed before. 

Finally, the \LayerDiagram of Figure~\ref{fig:mm:loop-sequence-layer} highlights that the \elem{:Self-management-1} module does not perform any adaptation activity since it has no label in contrast to the \elem{:Self-optimization} and \elem{:Self-repair} modules with their labels \elem{MAPE} and \elem{M..PE}. But the other example (cf. Figure~\ref{fig:mm:loop-overlapping-layer}) shows that the \elem{:Self-management-2} module performs the monitor and execute activities (\elem{M..E}) while the analyze and plan activities (\elem{AP}) are performed by the other modules.
Thus, an \LayerDiagram reflects all employed feedback loop instances and the modules coordinating the execution of these instances. This makes the coordinated execution visible at the architectural level.

\subsection{Modeling Layered Feedback Loops}\label{sec:ld:layered-architecture}

As argued in Section~\ref{sec:requirements}, there are particular cases, in which feedback loops are layered. A feedback loop at a higher layer adapts a feedback loop at the layer directly below. This requires adaptable feedback loops and appropriate reflections of these feedback loops.

Feedback loops specified by \EUREMA are adaptable by construction because \EUREMA models as feedback loop specifications and visualized by \FeedbackLoopDiagrams are kept alive at runtime and they are executed by an interpreter. The \EUREMA interpreter is able to cope with dynamic changes of \EUREMA models at runtime and even while executing these models. \EUREMA supports dynamic adaptation of feedback loops with respect to the \FeedbackLoopDiagram concepts. 
Thus, \EUREMA supports dynamically adjusting runtime models used within a feedback loop, for example, to replace the change models that define the planning. Model operations can be adjusted by adding, removing, or replacing them. This typically requires adapting the usage of runtime models and the control flow. Besides such structural adaptations, the control flow can be adjusted by parameter adaptation of conditions used for decision nodes.

Such adaptations of feedback loops are conducted by other feedback loops operating at higher layers. In \EUREMA, the modeling of higher-layer loops is similar to modeling any feedback loop as discussed in Section~\ref{sec:seams2012}.  
However, a particular aspect is the reflection model that represents the lower-layer loop and that is needed by the higher-layer loop for adaptation. Therefore, two variants corresponding to the idea of \emph{procedural} and \emph{declarative reflection} \cite{Maes1987} are supported, which will be discussed for the following example.

The self-adaptive software with its repair feedback loop as specified by the \FeedbackLoopDiagram in Figure~\ref{fig:mm:self-repair-withoutA} and the \LayerDiagram in Figure~\ref{fig:mm:self-repair-layer} automatically heals failures in the adaptable software by applying pre-defined repair strategies. However, these strategies need not to be able to handle \emph{all} failures since it is usually impossible to anticipate all kinds of failures when developing and deploying these strategies given the uncertainty in self-adaptive systems and their environments. Thus, the repair strategies defined in a runtime model have to be maintained and adapted at runtime. This task can be assigned to a higher-layer feedback loop that synthesizes new strategies on demand and provides them to the self-repair feedback loop.

\subsubsection{Procedural Reflection}\label{sec:ld:layered-architecture:procedural}

Keeping \EUREMA models alive at runtime, the models specifying and executing the lower-layer feedback loop can be directly used as a reflection model by the higher-layer loop. Thus, the higher-layer loop does not maintain a separate representation and one representation is used to specify, execute, and adapt the lower-layer~loop.

\begin{figure}[t]
\begin{minipage}[b]{0.5\linewidth}
\centerline{\includegraphics[scale=.264]{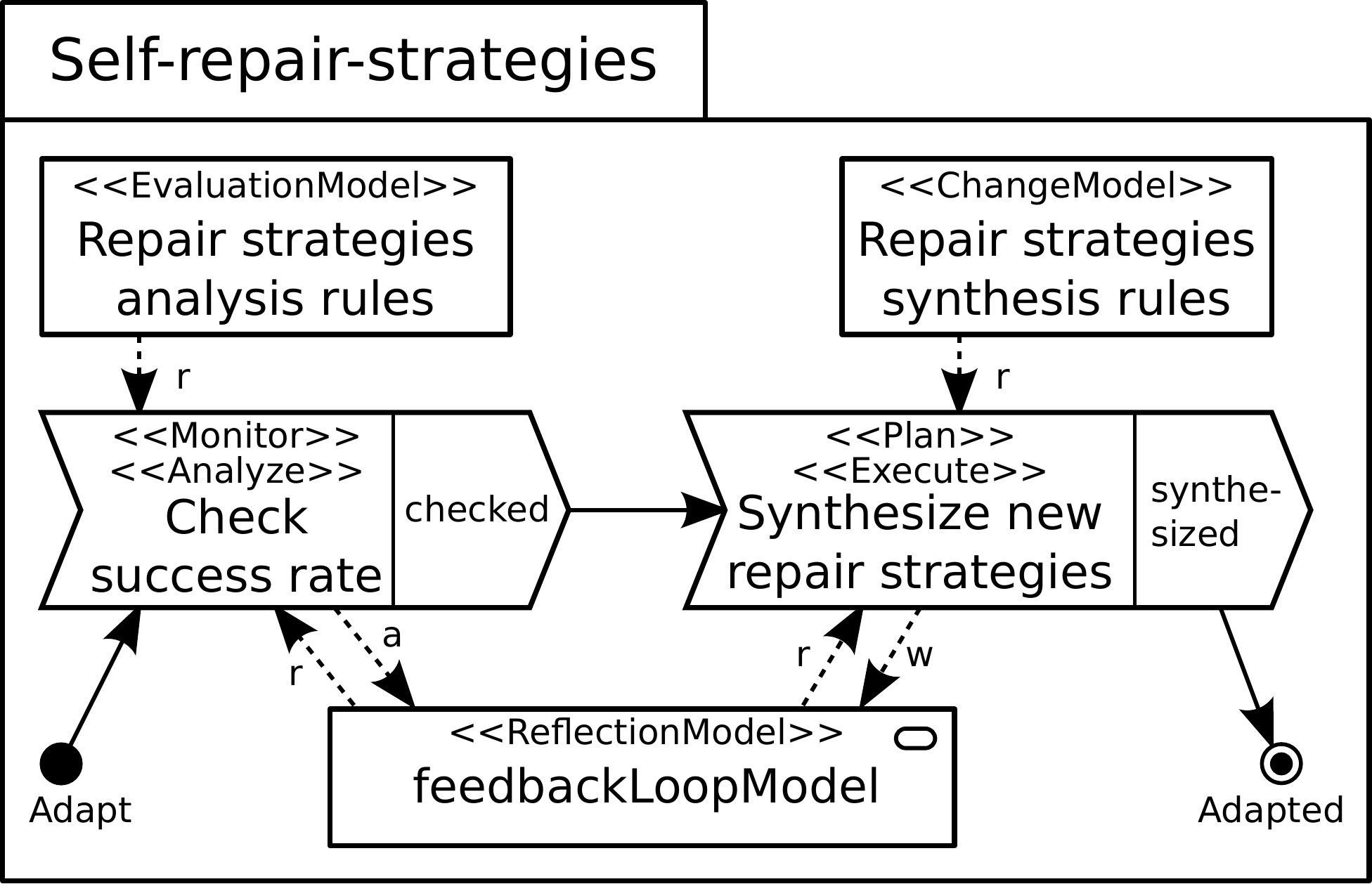}}
\caption{\FeedbackLoopDiagram for {\sf Self-repair-strategies}}
\label{fig:mm:self-repair-strategies}
\end{minipage}%
\begin{minipage}[b]{0.5\linewidth}
\centerline{\includegraphics[scale=.3]{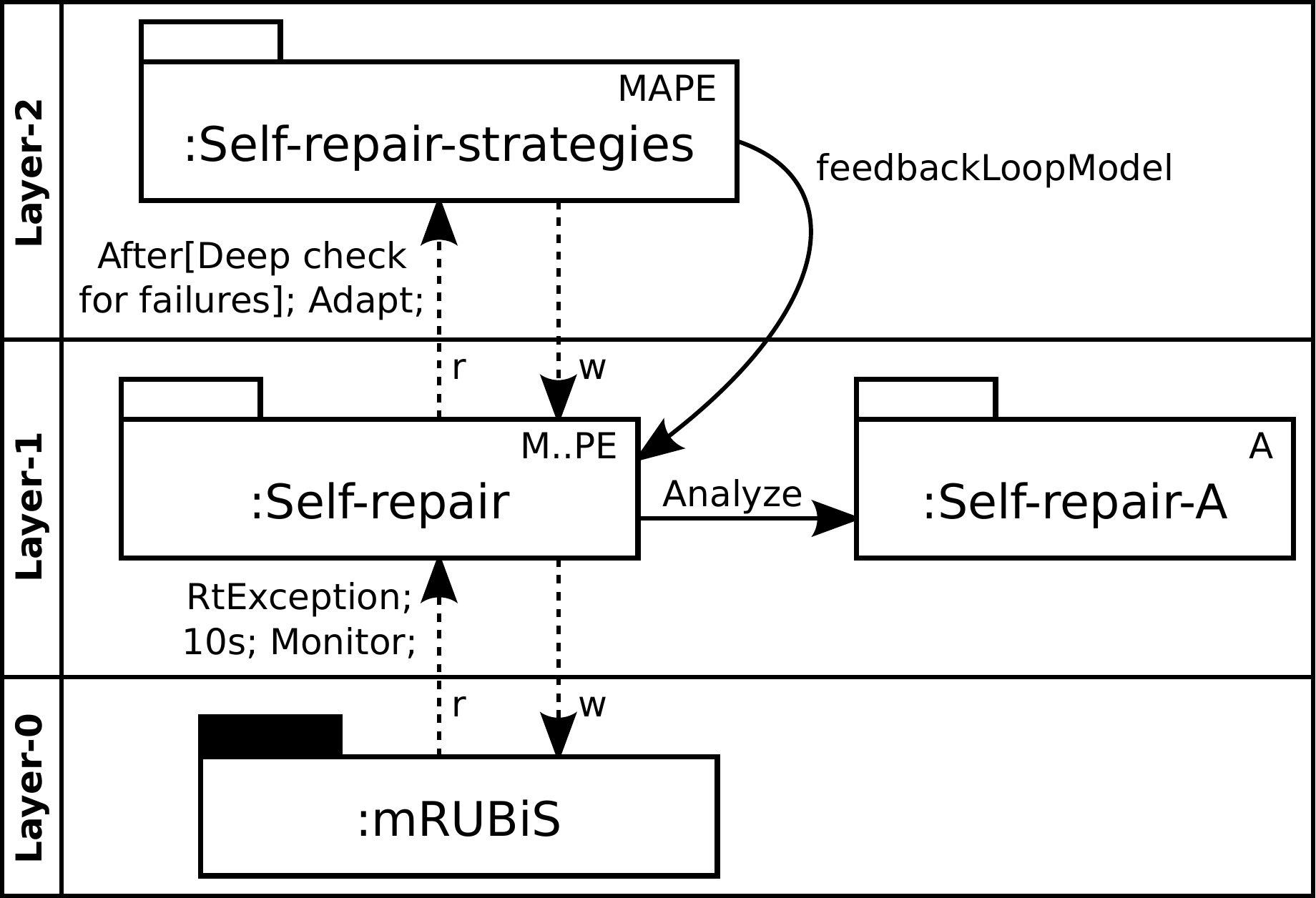}}
\caption{\LayerDiagram for {\sf Self-repair-strategies}}
\label{fig:mm:self-repair-strategies-layer}
\end{minipage}
\end{figure}

For the example, the corresponding higher-layer loop is defined by the \elem{Self-repair-strategies} \FeedbackLoopDiagram in Figure~\ref{fig:mm:self-repair-strategies}. It contains the reflection model \elem{feedbackLoopModel}, which is labeled with an icon to highlight that this model is directly an \EUREMA model as visualized by an \FeedbackLoopDiagram, particularly, the \FeedbackLoopDiagram that specifies and executes the self-repair loop (cf. Figure~\ref{fig:mm:self-repair-withoutA}). This model is used to check the success rate of the repair strategies and to synthesize new strategies that are directly provided to the self-repair loop by changing the model. Applying procedural reflection, the causal connection is ensured by construction and there is no need for explicit monitor and execute activities in the \elem{Self-repair-strategies} loop. These activities are implicitly realized by the analyze and plan activities (cf. related stereotypes in Figure~\ref{fig:mm:self-repair-strategies}). 

At runtime, the \elem{feedbackLoopModel} reflecting the self-repair feedback loop has to be bound to the specific \FeedbackLoopDiagram instance executing this loop. This is defined by the \LayerDiagram in Figure~\ref{fig:mm:self-repair-strategies-layer} showing the \elem{:Self-repair-strategies} module at \mbox{\elem{Layer-2}} that senses and effects the \elem{:Self-repair} module at \mbox{\elem{Layer-1}}. The binding of the \elem{feedbackLoopModel} to the specific \FeedbackLoopDiagram instance of the \elem{:Self-repair} module is defined by the use relationship having the name as the reflection model.
To adapt the self-repair loop, the \elem{:Self-repair-strategies} module operates on this \FeedbackLoopDiagram instance, whose specification is shown in Figure~\ref{fig:mm:self-repair-withoutA}. Thereby, it also operates on the instance of the \elem{Self-repair-A} \FeedbackLoopDiagram shown in Figure~\ref{fig:mm:self-repair-A} because this \FeedbackLoopDiagram is used and therefore included by the \elem{Self-repair} \FeedbackLoopDiagram.
In general, to dynamically adapt a lower-layer feedback loop, a higher-layer loop changes the \FeedbackLoopDiagram instance that specifies and executes the lower-layer loop. Thus, all \FeedbackLoopDiagrams shown in this article are \emph{initial} specifications of feedback loops since they are kept alive at runtime and they might be dynamically changed at runtime.

The triggering condition of the \elem{:Self-repair-strategies} module (cf.~Figure~\ref{fig:mm:self-repair-strategies-layer}) is similar to the conditions discussed in Section~\ref{sec:ld:trigger}. It refers to an event, namely \elem{After[Deep check for failures]}, emitted by the sensed modules, particularly by the \EUREMA interpreter when executing the \elem{:Self-repair} and \elem{:Self-repair-A} modules. In general, the interpreter synchronously emits two types of events when executing an \FeedbackLoopDiagram instance: \elem{Before[opName]} and \elem{After[opName]} events are emitted \emph{before} and respectively \emph{after} any model operation is executed while \elem{opName} refers to the name of the operation. This supports intercepting the execution of a (lower-layer) feedback loop to synchronously execute another (higher-layer) feedback loop.
In the example, the \elem{:Self-repair-strategies} module is triggered after executing the self-repair's \elem{Deep check for failures} operation. This is the case if more than five consecutive runs of the self-repair loop were not able to repair the failures, which indicates the need for new repair~strategies.

The advantage of directly using \EUREMA models as reflection models of feedback loops is that the causal connection is ensured by construction. This avoids the development of monitor and execute activities for higher-layer loops, which create and maintain reflection models of lower-layer loops. However, by using the same model of a loop for executing as well as adapting it, the adaptation cannot be decoupled from the execution. Thus, any adaptation performed by the higher-layer loop is instantaneously enacted to the lower-layer~loop.

\subsubsection{Declarative Reflection}\label{sec:ld:layered-architecture:declarative}

\begin{figure}[t]

\begin{minipage}[b]{0.5\linewidth}
\hspace{0.01\linewidth}
\centerline{\includegraphics[scale=.264]{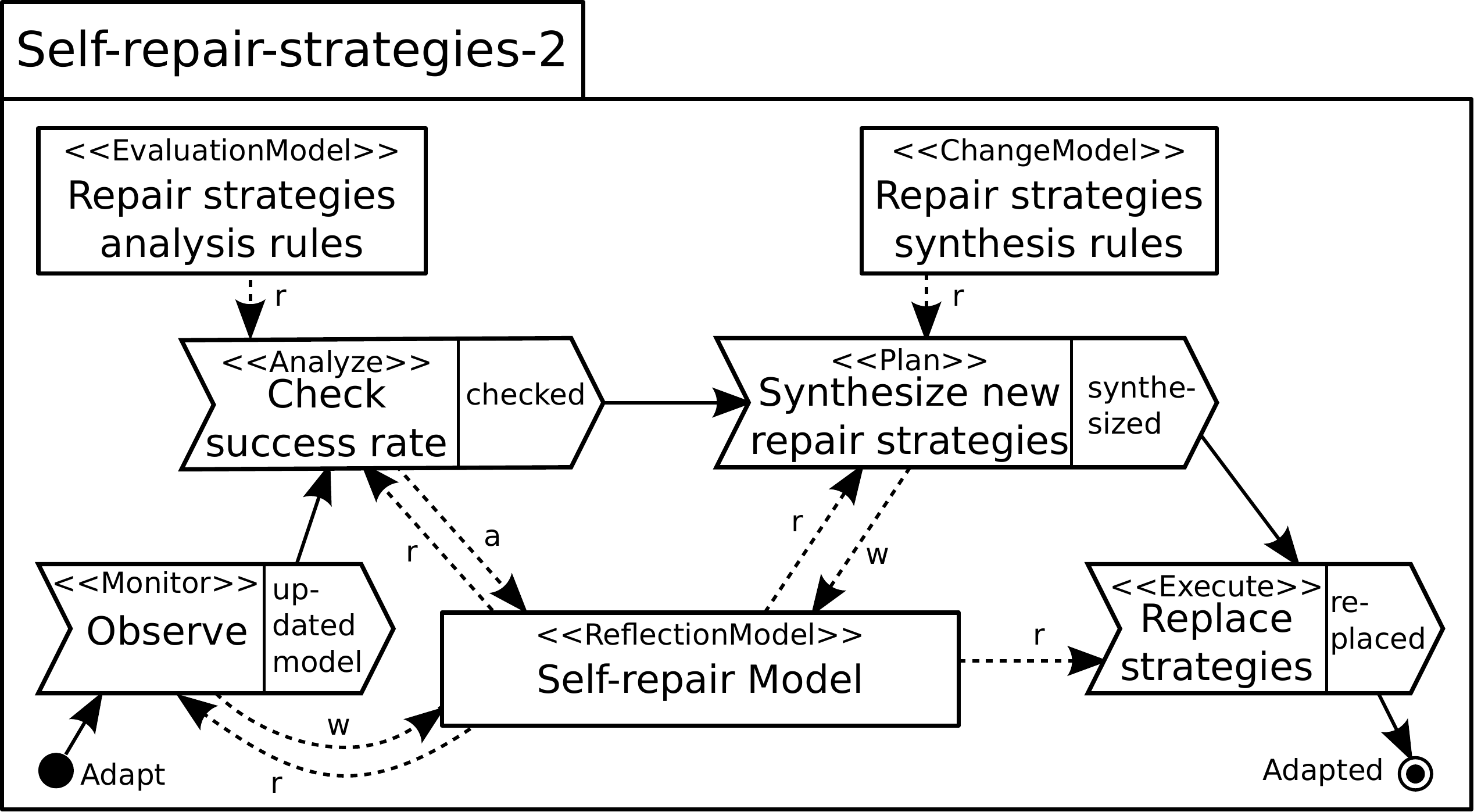}}
\caption{\FeedbackLoopDiagram for {\sf Self-repair-strategies-2}}
\label{fig:mm:self-repair-strategies-2}
\end{minipage}%
\begin{minipage}[b]{0.5\linewidth}
\centerline{\includegraphics[scale=.3]{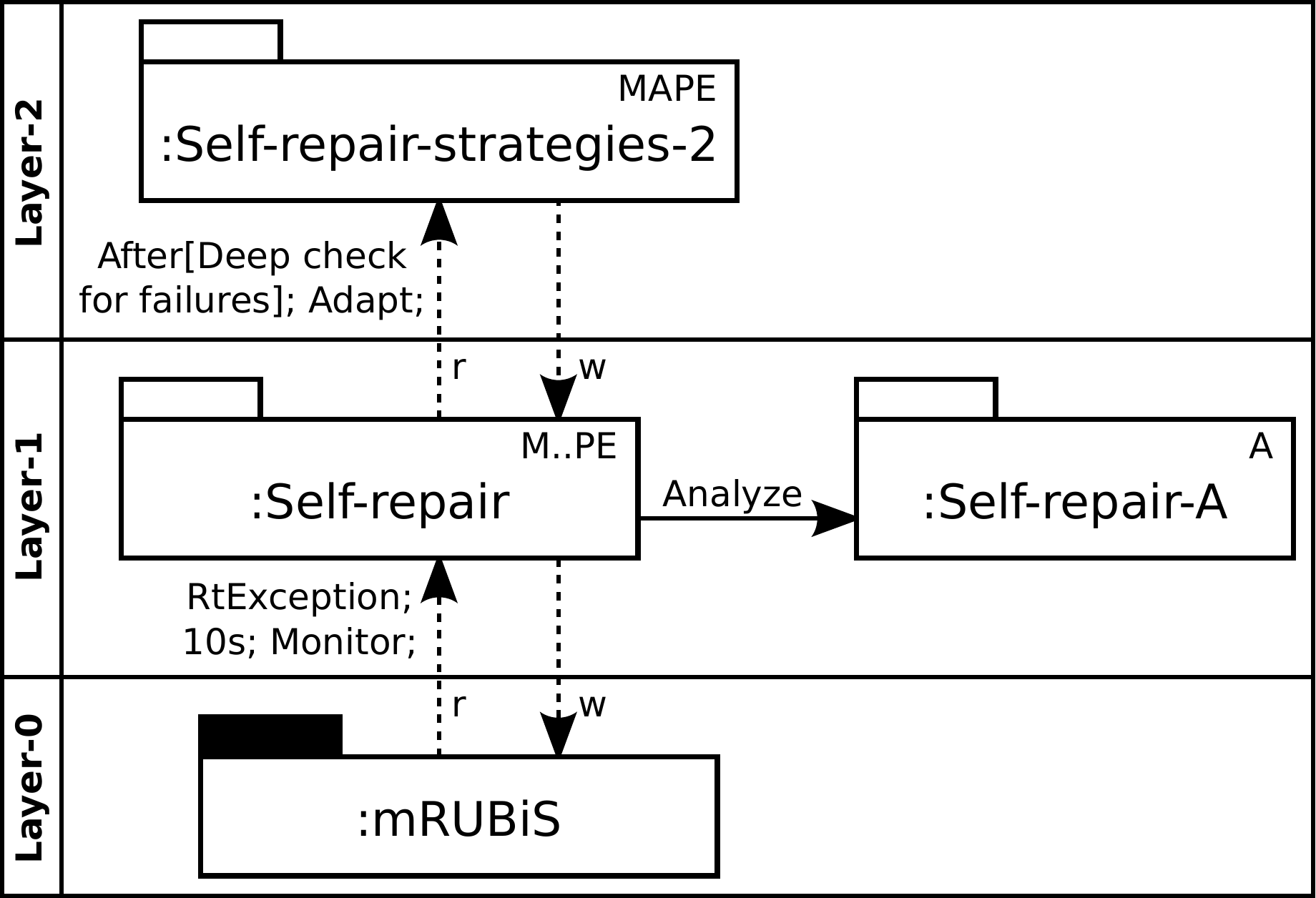}}
\caption{\LayerDiagram for {\sf Self-repair-strategies-2}}
\label{fig:mm:self-repair-strategies-2-layer}
\end{minipage}
\vspace{0.15em}
\end{figure}

In declarative reflection, a higher-layer feedback loop employs a user-defined reflection model of the lower-layer feedback loop. Thus, two representations of the lower-layer loop are maintained, one for specifying and executing it, and one for adapting it.
To realize the example we used for discussing procedural reflection, the higher-layer loop is specified by the \FeedbackLoopDiagram in Figure~\ref{fig:mm:self-repair-strategies-2}. The monitor activity observes the self-repair loop and maintains the \elem{Self-repair Model} that reflects the self-repair loop. Using this reflection model, the analyze activity checks the success rates of the current repair strategies. The plan activity synthesizes new strategies that replace the current ones in the reflection model. This replacement is enacted to the \elem{:Self-repair} module (instance of the \FeedbackLoopDiagram from Figure~\ref{fig:mm:self-repair-withoutA}) by the execute activity that replaces the \elem{Repair strategies} runtime model in this module. 

As defined by the \LayerDiagram in Figure~\ref{fig:mm:self-repair-strategies-2-layer}, an instance of the \elem{Self-repair-strategies-2} feedback loop at \elem{Layer-2} senses and effects the \elem{:Self-repair} feedback loop including the \elem{:Self-repair-A} activity at \elem{Layer-1}.
A particular aspect of this layering is that the higher-layer loop utilizes a user-defined reflection model (cf. \elem{Self-repair Model} in Figure~\ref{fig:mm:self-repair-strategies-2}) such that no binding of this model has to be specified in the \LayerDiagram. This model is user-defined because its metamodel can be user-defined and it is maintained by user-defined model operations for the monitor and execute activities. 
Thus, the engineer may decide which information about the lower-layer loop is covered by the reflection model as well as the abstraction level of the model. However, she must ensure the causal connection between the reflection model and the reflected feedback loop by defining and implementing model operations for the monitor and execute activities.

Therefore, sensors and effectors provided by \EUREMA can be used to observe and adjust feedback loops by means of the \EUREMA models. For sensing \EUREMA models, they can be queried and events notifying about the execution and changes of these models are emitted by the \EUREMA interpreter and the MDE infrastructure, in particular the \emph{Eclipse Modeling Framework} (EMF). For effecting \EUREMA models, basic means to change models are provided such as changing attribute values, or adding and removing nodes and relationships.

The advantage of user-defined reflection models is that the higher-layer feedback loop may run decoupled from the lower-layer loop since the reflection model is kept separate from the \EUREMA model specifying and executing the lower-layer loop. 
However, the disadvantage is that both representations have to be synchronized to each other to ensure the causal connection. Nevertheless, this synchronization can be simplified since both representations are models conforming to MDE principles, that is, they have potentially different metamodels but the same meta-metamodel. Thus, one-to-one copies of \EUREMA models can be directly provided as reflection models or MDE techniques such as model synchronization to keep both models consistent to each other can be employed. The applicability of such techniques for runtime reflection models has been shown in~\cite{VogelNHGB10,VG10}. 

Overall, \LayerDiagrams explicitly reflect the feedback loops at individual layers and whether procedural or declarative reflection is employed among them. If a reflection model is bound to a megamodel module (\FeedbackLoopDiagram instance) by a use relationship in the \LayerDiagram, procedural reflection is employed. If no such binding is defined, declarative reflection is employed.

\subsection{Off-line Adaptation}\label{sec:ld:offline-adaptation}

As motivated in Section~\ref{sec:requirements}, self-adaptive software must support the co-existence of on-line and off-line adaptation to ensure its long-term evolution. On-line adaptation refers to activities performed by the adaptation engine, and off-line adaptation to activities performed by engineers to maintain the software. 
Evolving self-adaptive software considers scenarios such as
(1)~adding or removing feedback loops from the running software, 
and changing 
(2)~running feedback loops,
(3)~legacy feedback loops, and
(4)~the running adaptable software.
The rationale of our approach to the co-existence is to interpret such evolution steps as feedback loops split up into on-line monitor and execute and off-line analyze and plan~activities.

We support on-line monitoring by observing and exporting snapshots of \EUREMA models including the runtime models that are used within the feedback loops. Since \EUREMA models as visualized by \FeedbackLoopDiagrams and \LayerDiagrams are directly used for executing \mbox{adaptation} engines, they innately reflect the running feedback loops and the run\-time architecture of the self-adaptive software. 
These snapshots are transferred to~the \mbox{development} environment and support the engineer in analyzing the software and planning an adaptation offline. Planning includes modeling the enactment of the adaptation with \EUREMA.
The resulting models are loaded up to the \EUREMA interpreter that dynamically instantiates and executes these models. This executes the adaptation on-line to the running software and accomplishes the evolution step.
A particular issue is to execute on-line such an adaptation while feedback loops are operating in the adaptation engine. This requires a coordinated execution, which is explicitly specified by \EUREMA models.
In the following, we discuss \EUREMA's support for the co-existence using the four evolution scenarios we just outlined above.

\subsubsection{Adding and Removing Feedback Loops}\label{sec:ld:offline-adaptation:addFLD}
Assuming the example self-adaptive software that only employs the self-repair feedback loop, we have discussed the need for a higher-layer feedback loop that maintains the repair strategies in Section~\ref{sec:ld:layered-architecture}. This need should be addressed by an off-line adaptation that equips the software with the higher-layer loop.

Observing and analyzing the running self-repair feedback loop by means of the \EUREMA models (\LayerDiagram in Figure~\ref{fig:mm:self-repair-layer} and instances of the \FeedbackLoopDiagrams from Figures~\ref{fig:mm:self-repair-A} and~\ref{fig:mm:self-repair-withoutA}), their execution, and the contained runtime models, an engineer identifies the need to maintain the repair strategies as failures are continuously identified. Having developed mechanisms to automatically synthesize repair strategies, the engineer may delegate the maintenance of the strategies to the adaptation engine by specifying offline a corresponding feedback loop with \EUREMA. This results in the \FeedbackLoopDiagram of Figure~\ref{fig:mm:self-repair-strategies} that was discussed in Section~\ref{sec:ld:layered-architecture}.

\begin{figure}
\centerline{\includegraphics[scale=.3]{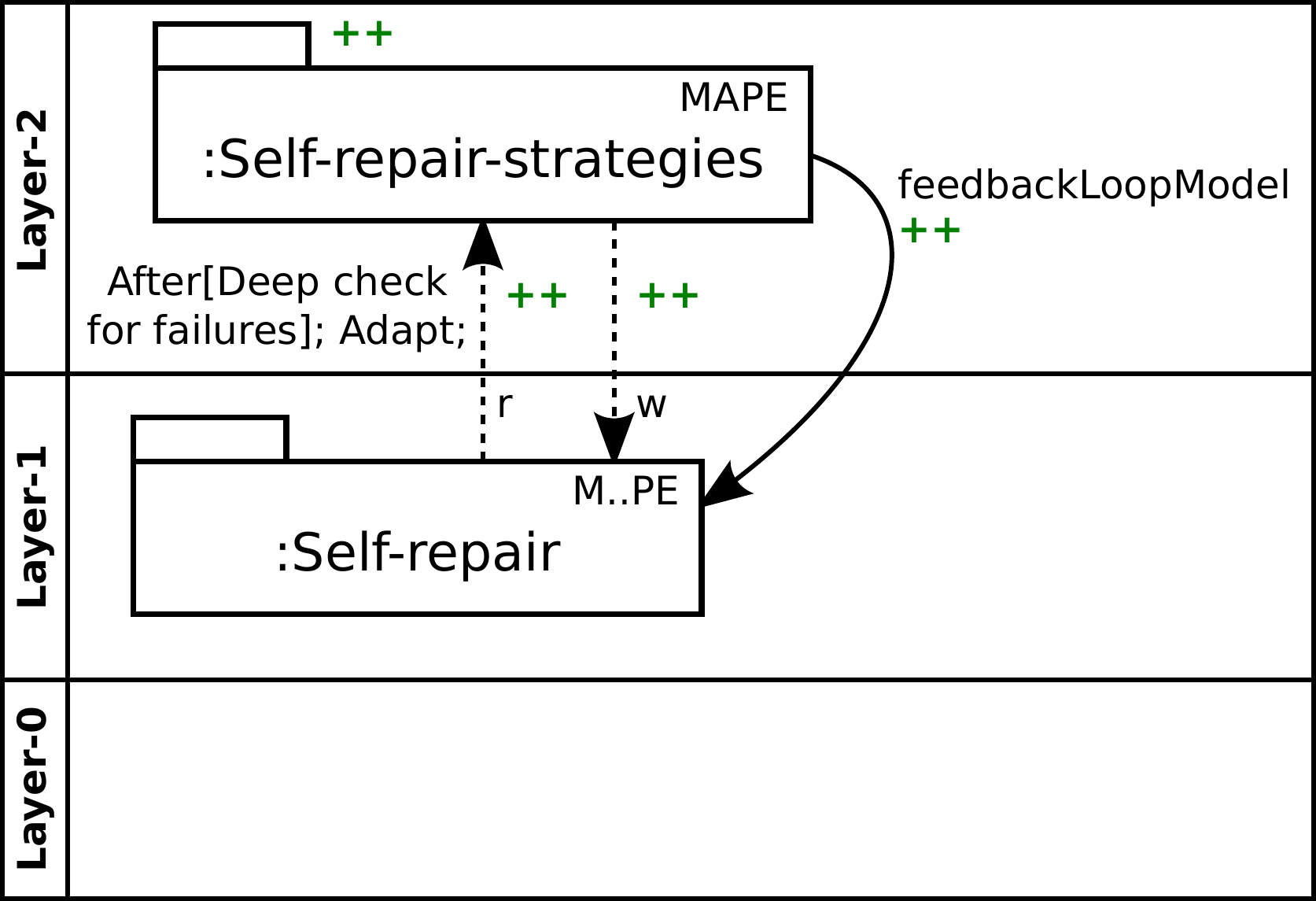}}
\caption{A rule integrating a module}
\label{fig:mm:self-repair-strategies-rule}
\vspace{.5em}
\end{figure}

The engineer loads up this \FeedbackLoopDiagram to the adaptation engine together with a rule defining how this \FeedbackLoopDiagram should be integrated as a megamodel module into the engine's architecture. The architecture is reflected by the \LayerDiagram in Figure~\ref{fig:mm:self-repair-layer} and the integration is defined by a graph transformation rule\footnote{\EUREMA does not define or prescribe the language for such integration rules. Theoretically, any mechanism can be used that specifies changes of a model. We apply a mechanism based on graph transformations.} (cf. Figure~\ref{fig:mm:self-repair-strategies-rule}) changing this \LayerDiagram. The rule's elements that are annotated with \elem{++} are added to the \LayerDiagram if a match for its elements having no annotation is found in the \LayerDiagram. Thus, the \EUREMA interpreter instantiates the uploaded \FeedbackLoopDiagram to the \elem{:Self-repair-strategies} module and applies the rule. 
This identifies a match for the \elem{:Self-repair} module at \elem{Layer-1}, adds the \elem{:Self-repair-strategies} module at \elem{Layer-2}, establishes the sense and effect relationships between both modules, and binds the \elem{feedbackLoopModel} used in the \elem{:Self-repair-strategies} module to the \elem{:Self-repair} module (cf. Section~\ref{sec:ld:layered-architecture}).
Overall, this results in a three-layer architecture reflected by the \LayerDiagram in Figure~\ref{fig:mm:self-repair-strategies-layer}.
While a module is integrated into the engine, the already \mbox{existing} modules may operate without any impact except of delays. Such delays are caused~by supporting consistent architectural reconfiguration in a quiescent state. Then, the execution of a new module is coordinated with the other modules by a triggering condition as defined in the integration rule (e.g., Figure~\ref{fig:mm:self-repair-strategies-rule}). As discussed in Section~\ref{sec:ld:layered-architecture}, such a trigger for a megamodel module that senses another megamodel module enables the exclusive execution by intercepting the execution of the lower-layer module to run the higher-layer module. 
This enables the exclusive and therefore synchronized execution of the modules to coordinate the online execution of an offline adaptation with running feedback loops.

In general, an \LayerDiagram is used at runtime as a procedural reflection model of the self-adaptive software, which can be dynamically changed to adjust the adaptation engine. This is accomplished in the context of off-line adaptation by applying an integration rule (cf. Figure~\ref{fig:mm:self-repair-strategies-rule}) while it is conceivable to employ another adaptation engine on top of the engine reflected by the \LayerDiagram.
The following dynamic changes of adaptation engines are supported. Layers and megamodel modules (\FeedbackLoopDiagrams instances) can be added and removed from the engine. An individual module can be adapted by changing the bindings between its complex model operations and other megamodel modules, and between its basic model operations and software modules implementing these operations. This has been discussed in Section~\ref{sec:ld:variability} in the context of variability that can now be exploited at runtime. These adaptations change the number and composition of modules in flexible layers of the engine. Finally, \LayerDiagrams reflect the triggering conditions of megamodel modules that can be dynamically changed.

\subsubsection{Changing a Running Feedback Loop}\label{sec:ld:offline-adaptation:changeFLD}
This scenario addresses changes of an already running feedback loop by an offline adaptation similar to a patch.
Thus, in contrast~to the previous scenario, the uploaded \FeedbackLoopDiagram does not specify a feedback loop but a patch process. For instance, an engineer has developed offline new repair strategies and she~specifies a patch process to just replace the strategies employed in the running self-repair feedback loop. The upload, integration, and execution of this \FeedbackLoopDiagram is identical to the previous scenario. The only difference is that the patch process is executed only once and then removed from the adaptation engine. This is accomplished by specifying a destruction state in the \FeedbackLoopDiagram (cf. Section~\ref{sec:seams2012:FLD}). 
This scenario exemplifies that by dynamic layers, feedback loops running in a certain layer can be dynamically changed by megamodel modules temporarily operating at the next higher layer to enact an off-line adaptation. Due to the similarities with the previous scenario, we refer to the technical report \cite{VG-TR13} for more details.

\subsubsection{Support for Adapting Legacy Feedback Loops}\label{sec:ld:offline-adaptation:legacy}
\EUREMA provides basic support for adapting legacy software modules realizing feedback loops. \EUREMA handles such modules as black boxes since they are not specified by \FeedbackLoopDiagrams and it just addresses their activation by reflecting them in \LayerDiagrams.
If it is possible to trigger legacy modules similar to \EUREMA feedback loops,  the \EUREMA interpreter can control their activation. Therefore, the interpreter can decommission legacy modules to migrate the adaptation engine to  megamodel modules specified by \EUREMA. This migration can be realized by off-line adaptation, especially to add and remove feedback loops (cf. Section~\ref{sec:ld:offline-adaptation:addFLD}). For more details on triggering legacy modules we refer to the technical report \cite{VG-TR13}.

\subsubsection{Changing the Adaptable Software}\label{sec:ld:offline-adaptation:adaptable-software}
Besides the adaptation engine, off-line adaptation may directly target the running adaptable software. For instance, a component of the adaptable software should be replaced as an improved version has been developed off-line due to maintenance requests. Since the employed self-repair feedback loop (cf. Figure~\ref{fig:mm:self-repair-layer}) is not able to replace a specific component, such an update has to be realized by another module.

\begin{figure}[t]
\centerline{\includegraphics[scale=.264]{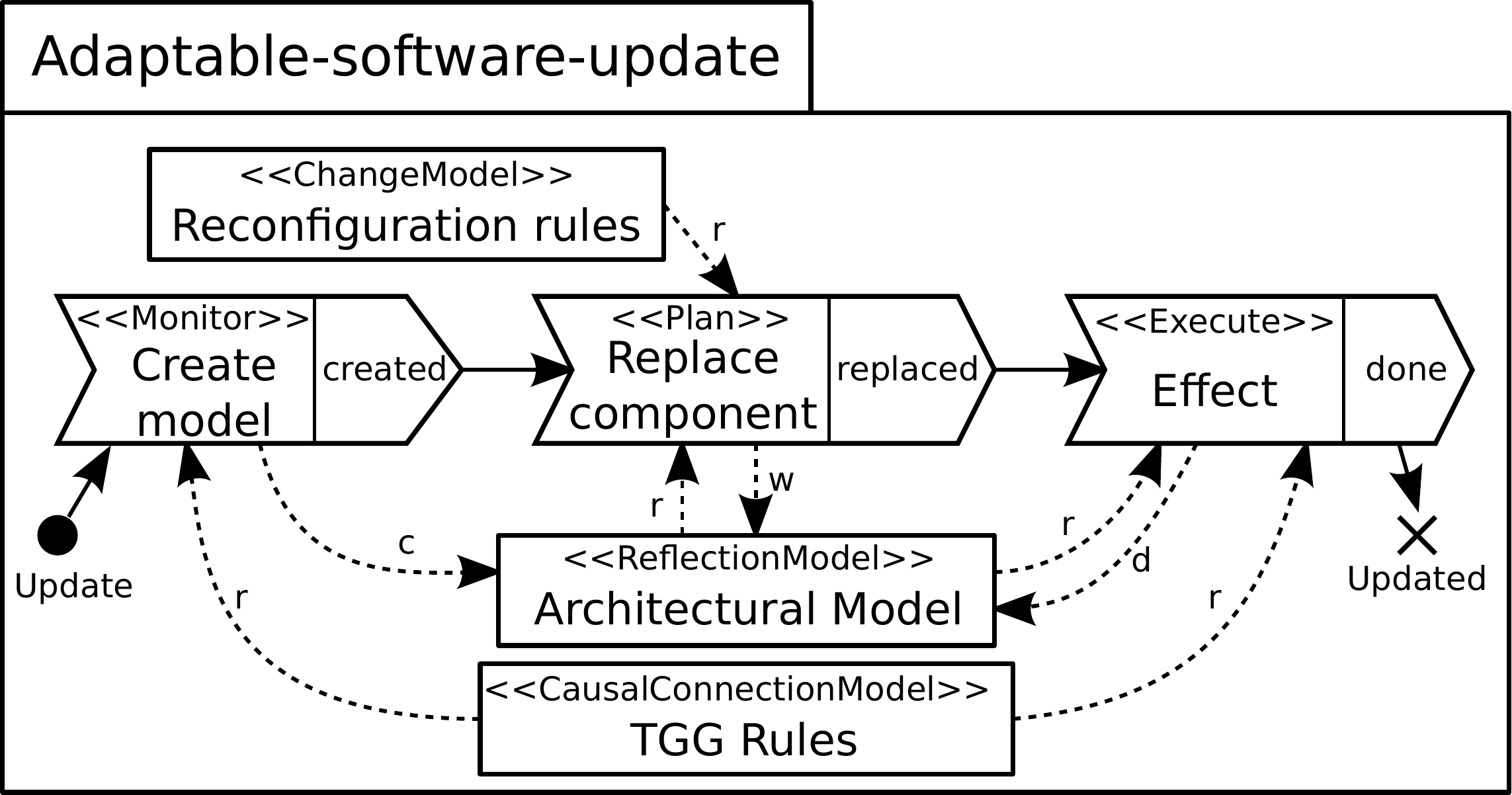}}
\caption{Updating the adaptable software}
\label{fig:mm:fld:adaptable-software-patch}
\end{figure}

Therefore, the engineer models the update by the \FeedbackLoopDiagram shown Figure~\ref{fig:mm:fld:adaptable-software-patch}. By monitoring, the \elem{Create model} operation creates the \elem{Architectural Model} reflecting the adaptable software. This model is used to reconfigure the architecture at the model level by applying rules defining the replacement of the specific component. The \elem{Effect} operation 
loads the new component, executes the reconfiguration prescribed in the model to the adaptable software, and finally destroys the architectural model.
Similar to the scenario that patches a feedback loop (cf. Section~\ref{sec:ld:offline-adaptation:changeFLD}), this \FeedbackLoopDiagram is loaded up, integrated, and executed once in the adaptation engine. Moreover, the execution is synchronized with the running self-repair feedback loop to avoid interferences while the adaptable software is reconfigured by the new module that is removed from the engine after the reconfiguration.
As discussed for the different scenarios, by reflection, dynamic layers, and dynamically (un)loading of \FeedbackLoopDiagram, \EUREMA supports off-line adaptation to evolve self-adaptive software, particularly in a way that has not been anticipated when initially deploying the software.

\section{Metamodel and Interpreter Implementation}\label{sec:implementation}

While the \EUREMA metamodel and its execution semantics are discussed in 
our technical report \cite{VG-TR13} due to space constraints, we briefly discuss the implementation of the  metamodel and interpreter in this section.

Both have been developed with the Java-based \emph{Eclipse Modeling Framework}~(EMF). The stereotypes and labels of \FeedbackLoopDiagram and \LayerDiagram elements are not directly supported by the metamodel since they do not influence the execution semantics and therefore the \EUREMA interpreter. However, they are introduced in \EUREMA by a basic profile mechanism to use them in the \EUREMA editor when modeling \FeedbackLoopDiagrams and \LayerDiagrams. The language for expressing conditions to exclusively branch the control flow in \FeedbackLoopDiagrams is defined by a grammar, implemented with the \emph{Java Compiler Compiler}~(JavaCC), and embedded in \EUREMA.

Our interpreter implementation only relies on EMF and it may run standalone and decoupled from the Eclipse workbench. Nevertheless, the interpreter provides full support for user-defined, EMF-based runtime models used within feedback loops. For example, the interpreter manages the handling of runtime models as input or output of model operation executions. Moreover, \EUREMA models as well as user-defined runtime models can be dynamically (un)loaded and exported as snapshots for off-line use. 
Concerning \EUREMA models, the interpreter currently provides full support for executing \FeedbackLoopDiagrams and almost full support for the \LayerDiagram. So far, multiple and layered feedback loops as defined in an \LayerDiagram can be executed but the \LayerDiagram is only implicitly maintained to support off-line adaptation~\cite{Hanysz2013}. Currently, we are addressing this aspect by making the \LayerDiagram an explicit runtime model.

\section{Discussion of Design Decision and Requirements Coverage}\label{sec:discussion}
In this section, we discuss \EUREMA by means of its fundamental design decisions that lead to the presented language and its coverage of the requirements identified in Section~\ref{sec:requirements}.

\subsection{Design Decisions}\label{sec:discussion:design-decisions}
The language design of \EUREMA is motivated by the requirements for engineering self-adaptive software (cf. Section~\ref{sec:requirements}).
With the \FeedbackLoopDiagram, we aim for an explicit specification of feedback loops (\textbf{R1}). According to the MAPE-K blueprint, a feedback loop consists of adaptation activities sharing knowledge. 
This calls for language concepts that describe these activities and the knowledge that we refined to a set of runtime models. 
This further requires concepts that make runtime models first class citizens of the models that specify feedback loops. Therefore, we applied generic megamodel concepts to the language design of \FeedbackLoopDiagrams. A megamodel refers to a model that contains other models and relationships between these models. The relationships can be substantiated to operations that manipulate these models \cite{BDDFB07,Bezivin_et_al:2003,Bezivin+2004,favre:DSP:2005:13}. Thus, models are first class elements of a megamodel. Consequently, we consider an \FeedbackLoopDiagram as a megamodel describing the adaptation activities as operations that work on runtime models (\textbf{R7}).
Hence, \emph{operations} and \emph{runtime models} are the main language concepts of \FeedbackLoopDiagrams. To make the \FeedbackLoopDiagrams executable (\textbf{R6}) and to address the intra-loop coordination (\textbf{R2}), we extended these concepts with the \emph{control flow} among the activities and the \emph{usage of models} by the activities. This explicitly defines the workflow of activities and the data that is consumed or produced by these activities. Using these concepts, an individual feedback loop can be specified by \FeedbackLoopDiagrams.
Moreover, these concepts are sufficient to specify multiple feedback loops and their coordinated execution such that no further concepts are required for inter-loop coordination (\textbf{R4}).
Finally, we designed \FeedbackLoopDiagrams to be runtime models. Keeping feedback loop specifications alive at runtime and interpreting them leverage adaptable feedback loops in layered architectures (\textbf{R9}) and the provision of reflection models for such loops. This promises flexible adaptation engines.

While the \FeedbackLoopDiagrams address the behavioral specification of feedback loops, they do not provide a structural view of the self-adaptive software. 
For instance, our initial idea~for \EUREMA's approach to layered architectures as presented in \cite{VG12b} was only based on \FeedbackLoopDiagrams. Therefore, it had the following drawbacks. First, it only considered the triggering of a higher-layer feedback loop by an explicit invocation from the lower-layer loop. This required the anticipation of the higher-layer loop in the lower-layer loop. Second, it did not make the individual layers of the adaptation engine explicit such that it was not visible which feedback loop is located at which layer or which loop adapts which other loop. Third, it just considered procedural reflection among feedback loops. To address these issues, we substantially modified \EUREMA by introducing \LayerDiagrams that complement the behavioral view provided by \FeedbackLoopDiagrams. 
The \LayerDiagram concepts are motivated by layered architectures for self-adaptive software with multiple feedback loops.
Therefore, we introduced the \emph{layers} that contain \emph{modules}, particularly instances of feedback loops as specified by \FeedbackLoopDiagrams. Furthermore, to be open for adaptation behavior specified by other languages, we extended the concepts of modules to any software component such as legacy adaptation activities. Moreover, we introduced operational \emph{relationships} among modules to describe dependencies and to support the execution. These are the \emph{use} relationship to enable modular modules and their usage by each other, and the \emph{sense} and \emph{effect} relationships. The latter two make explicit which modules control which other modules. Such relationships are needed to reason about dependencies such as interferences between feedback loops.
In this context, an \LayerDiagram also constitutes a megamodel that primarily contains \FeedbackLoopDiagram instances and relationships among them.  

Besides specifying a layered architecture, we designed the \LayerDiagram to be a runtime model that can be used as a procedural reflection model of the adaptation engine. Therefore, the \LayerDiagram provides an instance view of the self-adaptive software that can be adjusted at~runtime. The goal of this decision was to facilitate off-line adaptation (\textbf{R10}). 
Finally, the triggering conditions (\textbf{R3}) for feedback loops are defined in the \LayerDiagram because they usually depend on other modules or require coordination with other modules. Thus, we decided not to define the triggering conditions in the \FeedbackLoopDiagrams that typically consider individual feedback loops. 

The proposed \EUREMA modeling language as used in \FeedbackLoopDiagrams and  \LayerDiagrams is specific for engineering feedback loops, which makes \EUREMA a domain-specific modeling language (DSML).
We propose a DSML instead of using an existing general-purpose software modeling language such as UML \cite{UML-Superstructure241} to clearly separate the adaptation logic (adaptation engine and feedback loops) from the domain logic (adaptable software). This avoids intertwining of the adaptation and domain logic and promotes separation of concerns as proposed by the external approach (cf. Section~\ref{sec:requirements}), which promises more maintainable and reusable designs \cite{Salehie&Tahvildari2009}.
Nevertheless, \FeedbackLoopDiagrams share concepts with flowcharts and data flow diagrams such as \emph{UML Activities}. \FeedbackLoopDiagrams and UML Activities are similar with respect to modeling flows of actions (in UML) or operations (in \EUREMA). However, in contrast to \EUREMA, UML does not provide megamodel concepts as first class entities, like a model being itself an element in another model.
Likewise, the \LayerDiagram borrows concepts from UML, particularly \emph{UML Packages} and \emph{UML Objects}. However, \LayerDiagrams focus on domain-specific concepts and elements such as layers, megamodel and software modules, and operational use, sense, and effect relationships among modules, which clearly separates the adaptation logic from the domain logic.
Additionally, we propose a lean metamodel for \EUREMA (cf. \cite{VG-TR13}) resulting in a light-weight and efficient interpreter that probably cannot be easily achieved for the complex UML metamodel.

With \EUREMA we adopt an MDE approach to leverage benefits of MDE to the \mbox{runtime}~environment as generally discussed by \citeN{France+Rumpe2008}. On the one hand, \EUREMA exploits MDE principles by means of its executable modeling language and interpreter. 
On the other hand, \EUREMA makes the feedback loop's knowledge explicit by runtime models. This additionally leverages MDE techniques at runtime to perform individual adaptation activities of a feedback loop. As \EUREMA is designed to target a reasonable abstraction level similar to the level of software architectures, adaptation activities are considered as abstract model operations. This enables the integration and reuse of existing MDE techniques and implementations for realizing and performing such activities.

For example, we employed an existing model synchronization engine for the monitor and execute activities to maintain an architectural runtime model of the adaptable software, and an \emph{Object Constraint Language} (OCL) engine for the analyze activity to check architectural constraints on this model \cite{VogelNHGB10,VG10}. Such engines can be considered as reusable implementations for adaptation activities.
This exemplifies that the development efforts for adaptation activities can be reduced by integrating and reusing existing engines from MDE. 
Moreover, since such engines are generic and they completely externalize the user-defined inputs in models such as OCL expressions, these models become runtime models. \EUREMA makes such runtime models explicit in \FeedbackLoopDiagrams and amenable for adaptation. For example, the OCL expressions can be dynamically adapted without having to change the OCL engine. This potentially simplifies the development of adaptable feedback loops.
Thus, \EUREMA directly exploits MDE principles for specifying, executing, and adapting feedback loops while it enables engineers to exploit MDE principles for implementing individual adaptation activities that are modeled as operations in \FeedbackLoopDiagrams.

We designed \EUREMA to be reusable such that it abstracts from sensor and effector details of the adaptable software. This avoids its coupling to the technology, platform, or type of a specific software. In contrast, \EUREMA proposes the explicit modeling of monitor and execute activities, whose implementations have to cope with these details. Sensors details are only revealed by sensor events that are used in triggering conditions of feedback loops.
Thus, we require that appropriate sensors, effectors, and activity implementations are available to realize parameter or structural adaptation (\textbf{R8}). \EUREMA then supports the modeling and coordinated execution of the activities in a feedback loop.

\subsection{Requirements Coverage}\label{sec:requirements-coverage}

As just discussed, the design of \EUREMA is motivated by the requirements for engineering self-adaptive software identified in Section~\ref{sec:requirements}. These requirements reflect the state-of-the-art issues that are discussed for self-adaptive software in research and \EUREMA provides concepts for almost all of them.
\FeedbackLoopDiagrams cover the explicit modeling of individual or multiple feedback loops (\textbf{R1}), their intra-loop (\textbf{R2}) and inter-loop coordination (\textbf{R4}), and their, however, non-concurrent execution (\textbf{R6}) (cf. Section~\ref{sec:seams2012}).
Thereby, \FeedbackLoopDiagrams can capture arbitrary runtime models and their usage in feedback loops without restricting the kinds of models by means of their metamodels and purpose (\textbf{R7}).
An \LayerDiagram covers the layered architecture of an instance of the self-adaptive software, which consists of all employed \FeedbackLoopDiagram instances, their triggering (\textbf{R3}), and their relationships to each other and to the adaptable software (cf.~Section~\ref{sec:ld}). Besides \FeedbackLoopDiagrams, this further makes the feedback loops visible in the design of self-adaptive software (\textbf{R1}) and shifts the abstraction level of feedback loops and their coordinated execution to the \mbox{architectural} level.
Moreover, by \LayerDiagrams and keeping executable \FeedbackLoopDiagram instances alive at runtime, adaptive or hierarchical control schemes in layered architectures are enabled (\textbf{R9}) (cf. Section~\ref{sec:ld:layered-architecture}).
Additionally, keeping the \LayerDiagram alive at runtime enables dynamic layered architectures, which supports off-line adaptation for the long-term evolution of self-adaptive software (\textbf{R10}) (cf.~Section~\ref{sec:ld:offline-adaptation}). 
Finally, \EUREMA addresses parameter and structural adaptation (\textbf{R8}) if corresponding sensors, effectors, and implementations of adaptation activities are available. Then, \EUREMA supports the modeling and coordinated execution of these activities. 

Overall, \EUREMA covers all of the requirements except of distribution (\textbf{R5}) and the entirely concurrent execution (\textbf{R6}). Thus, adaptation engines cannot be distributed and interdependent feedback loops and their adaptation activities cannot be executed concurrently.
Nevertheless, \EUREMA provides concepts for almost all issues as reflected by the requirements and currently discussed in the research literature on self-adaptive software such that it addresses a wide range of state-of-the-art problems for self-adaptive software.

\vspace{0.5em}
\section{Evaluation}\label{sec:evaluation}

Besides experimenting with \emph{mRUBiS} \cite{mRUBiS} as an application example, we evaluate \EUREMA by applying its language to state-of-the-art approaches to self-adaptive software and by investigating the runtime characteristics of its interpreter. This demonstrates the expressiveness of the language and the runtime efficiency of the interpreter, respectively.
 
\subsection{Application of the \EUREMA Language}\label{subsec:application}

\begin{figure}[t]
	\begin{minipage}[b]{0.35\linewidth}
		\centerline{\includegraphics[scale=.3]{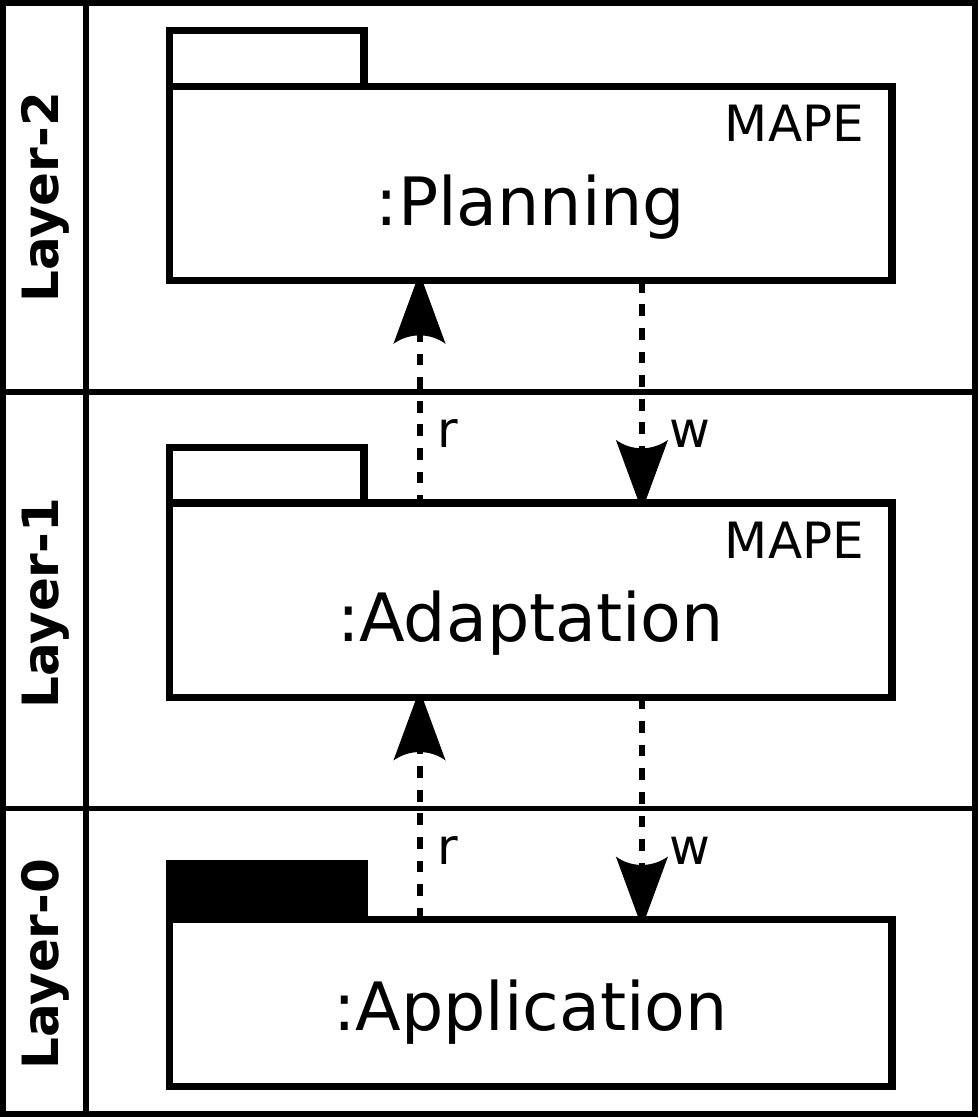}}
		\caption{\LayerDiagram for PLASMA}
		\label{fig:application:plasma:LD}
	\end{minipage}%
	\begin{minipage}[b]{0.65\linewidth}
		\centerline{\includegraphics[scale=.264]{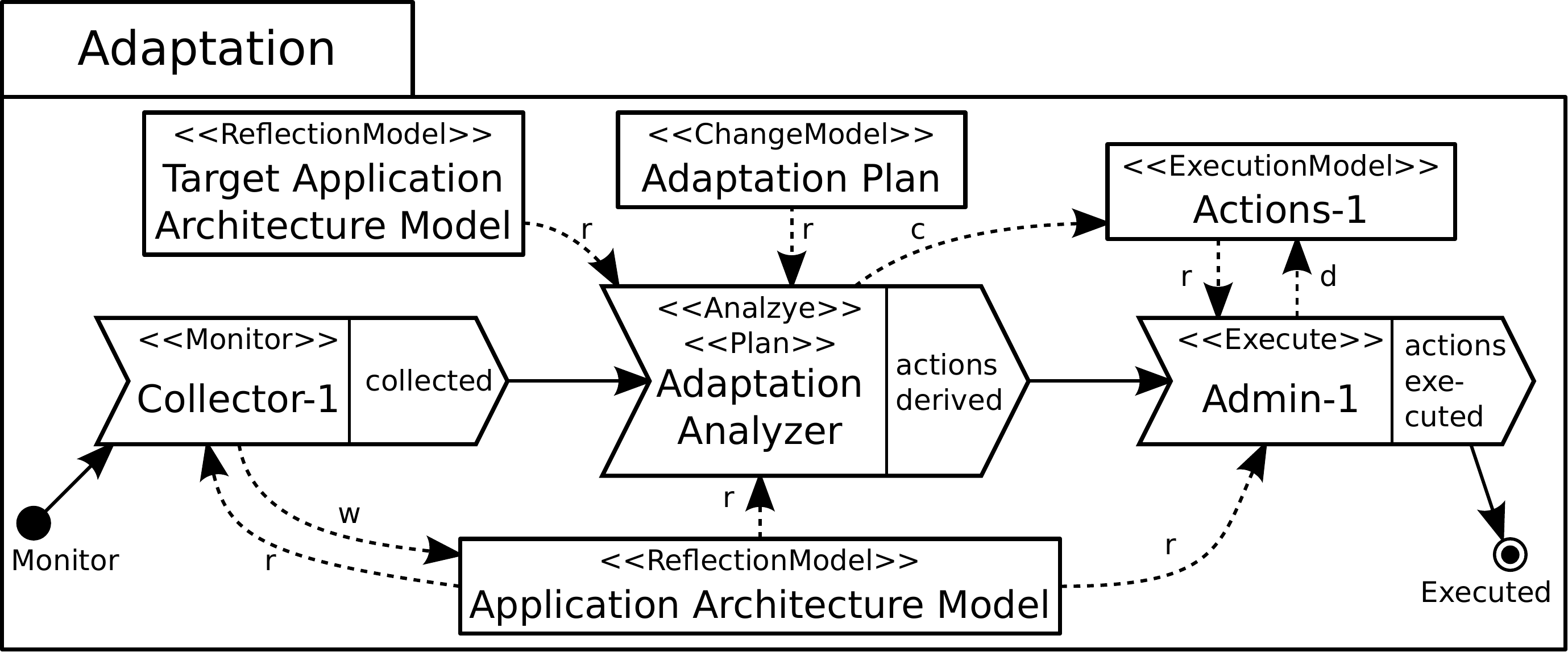}}
		\caption{\FeedbackLoopDiagram for the {\sf Adaptation} layer in PLASMA}
		\label{fig:application:plasma-adaptation:FLD}
	\end{minipage}
\vspace{0.5em}
\end{figure}

In this section, we investigate the expressiveness of the \EUREMA language by examples. Therefore, we applied it to state-of-the-art approaches from  literature, namely, \emph{Rainbow} \cite{GarCHSS04}, \emph{DiVA} \cite{Morin+2009}, and \emph{PLASMA} \cite{Tajalli2010}.
In this article, we will discuss the application to PLASMA that proposes a three-layer architecture for plan-based adaptation. This resulted in the following \EUREMA diagrams. 
The \LayerDiagram in Figure~\ref{fig:application:plasma:LD} defines the layered architecture with the adaptable application at the lowest layer. The feedback loop in the middle layer adapts the application and the highest-layer feedback loop (re)generates plans to be executed by the two lower layers.

The middle-layer loop is defined by the \FeedbackLoopDiagram in Figure~\ref{fig:application:plasma-adaptation:FLD}. The \elem{\mbox{Collector-1}} operation monitors the application and maintains the \elem{Application Architecture Model} reflecting the application. This model is used by the \elem{Adaptation Analyzer} to execute the \elem{Adaptation Plan} provided by the higher-layer loop. This plan specifies the adaptation to move the current application architecture to the target architecture defined in the \elem{Target Application Architecture Model}. Additionally, the \elem{Adaptation Analyzer} analyzes any deviations in the current application architecture and resolves them to align it with the target architecture. Therefore, reconfiguration commands (\elem{Actions-1}) are created and executed by the \elem{Admin-1} operation on the application.

\begin{figure}[t]
\centerline{\includegraphics[scale=.264]{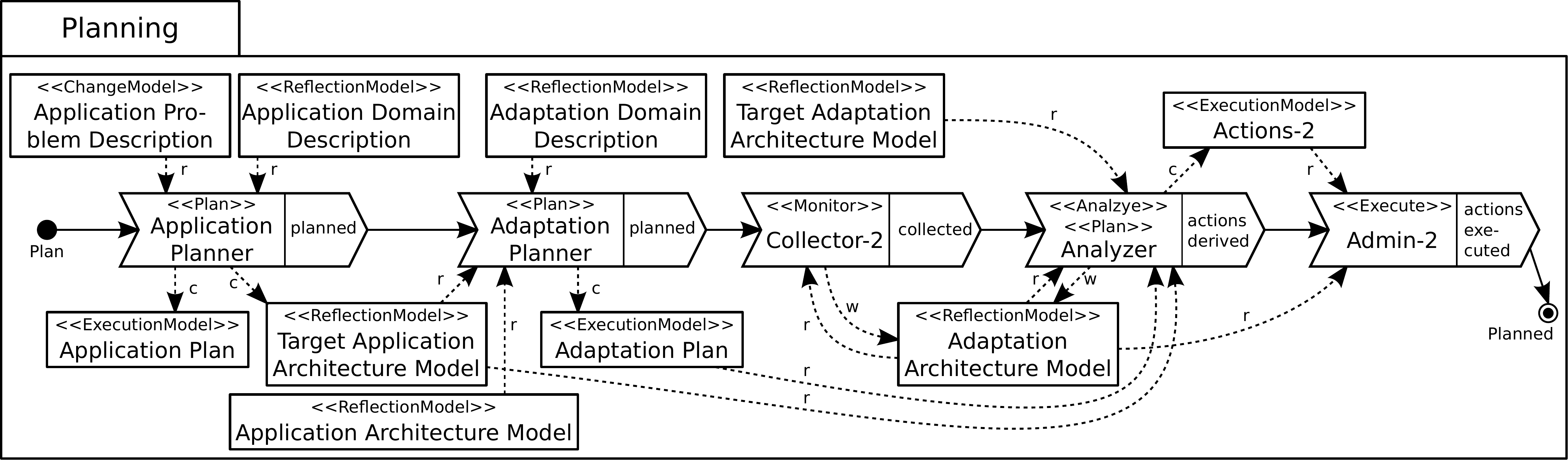}}
\caption{\FeedbackLoopDiagram for the {\sf Planning} layer in PLASMA}
\label{fig:application:plasma-planning:FLD}
\vspace{-.5em}
\end{figure}

The higher-layer feedback loop defined by the \FeedbackLoopDiagram in Figure~\ref{fig:application:plasma-planning:FLD} is executed when plans are generated initially or when replanning is required. 
The \elem{Application Planner} uses a domain model of the application (\elem{Application Domain Description}) and the initial and goal states of the application (\elem{Application Problem Description}), which are all provided by the architect. 
The planner creates an \elem{Application Plan} to be executed by the application at the lowest layer and the \elem{Target Application Architecture Model} prescribing the application architecture that is able to execute the application plan. 
Moreover, the architect provides the \elem{Target Adaptation Architecture Model} defining the target architecture of the middle-layer feedback loop, which is also reflected in the \elem{Adaptation Domain Description} used by the \elem{Adaptation Planner}. This planner additionally uses the \elem{Application Architecture Model} maintained by the middle-layer feedback loop and the newly created \elem{Target Appli\-cation Architecture Model} to derive an \elem{Adaptation Plan} defining how to move the current ar\-chitecture of the application to the target architecture. 
Then, the~\mbox{following}~\mbox{operations} adapt the middle-layer loop to enable the execution of the generated adaptation plan. The \elem{\mbox{Collector-2}} operation updates the \elem{Adaptation Architecture Model} reflecting the middle-layer feedback loop by monitoring. The \elem{Analyzer} adds the \elem{Adaptation Plan} and the \elem{Target Application Architecture Model} to this model in order to provide them to the middle-layer loop. Moreover, based on the current and target architectures of the middle-layer loop, reconfiguration commands (\elem{Action-2}) are generated to adapt this loop, for example, to replace the \elem{Adaptation Analyzer} (cf.~Figure~\ref{fig:application:plasma-adaptation:FLD}) with a version that is able to execute the new adaptation plan. Finally, the \elem{Admin-2} operation adapts the middle-layer loop by executing the reconfiguration commands and synchronizing the \elem{Adaptation Architecture Model}.

Overall, the \EUREMA language is able to capture PLASMA's architecture and feedback loops. However, the proper modeling of PLASMA is hard to assess since PLASMA only implicitly maintains the runtime models in a middleware and it does not make their usage by the loops and the triggers of the loops explicit. Thus, we derived as far as possible the runtime models and their usage from \cite{Tajalli2010} but we omitted the triggers. 

Besides PLASMA, we applied the \EUREMA language to Rainbow and DiVA, which is discussed in our technical report \cite{VG-TR13}.
The results of modeling these three examples demonstrate that the \EUREMA language is expressive enough to capture these state-of-the-art approaches and different variants of feedback loops. These variants especially concern the techniques that drive the feedback loops (architecture description languages in PLASMA and Rainbow, and MDE in DiVA) and the number of feedback loops (two layered feedback loops in PLASMA, and single feedback loops in Rainbow and DiVA). 
Though these examples provide only limited evidence for the expressiveness of  the \EUREMA language, we have already shown at the conceptual level that \EUREMA is expressive enough to addresses a wide range of state-of-the-art problems for self-adaptive software (cf. Section~\ref{sec:requirements-coverage}).

\subsection{Runtime Characteristics of the \EUREMA Interpreter}\label{sec:execution-results}

Finally, we evaluate the \EUREMA interpreter by discussing its runtime characteristics. We conducted experiments to quantify the load and overhead of the interpreter compared to a code-based solution to execute the self-repair feedback loop defined by the \FeedbackLoopDiagrams in Figures~\ref{fig:mm:self-repair-A} and~\ref{fig:mm:self-repair-withoutA}.
For the experiments, we considered the case, in which each run of the feedback loop instance always identifies failures. Moreover, a warm-up phase taking place before the actual measurements executes the instance more than five times, such that the condition branching the control flow in the \FeedbackLoopDiagram depicted in Figure~\ref{fig:mm:self-repair-A} is always fulfilled. Thus, for the measurements, all of the five basic model operations of the self-repair feedback loop are executed in each run of the instance.
As implementations for these model operations, we provided software modules as mocks that have runtime models as input as it is defined in the \FeedbackLoopDiagrams. Moreover, all runtime models that are the output of any model operation are already used as input of the same operation. Thus, no new models are produced by the mocks. In contrast, all runtime models are pre-defined and they are not changed at all by the mocks. Each mock can be assigned a duration, for which it generates load to simulate computations of the model operations. 

To evaluate the runtime characteristics of the \EUREMA interpreter, we implemented a code-based solution in \emph{Java} that executes the self-repair feedback loop. This solution does not use any \EUREMA model but it hard-codes  the execution by sequentially invoking the five mocks, one for each model operation of the self-repair feedback loop. Moreover, this code-based solution provides the runtime models required as input for invoking the mocks.
   
The experiments are configured by two parameters.
First, the duration assigned to the mocks defines the internal computation time of the model operations. The same duration is assigned to all mocks for one experiment and they vary for the different experiments. This results in four groups of experiments, either assigning a duration of 0ms, 5ms, 10ms, or 20ms to each mock. Since the self-repair feedback loop has five basic model operations, this constitutes a total computation time of either 0ms, 25ms, 50ms, or 100ms for one run of the feedback loop instance.   
The second parameter is the frequency of consecutive runs of the instance, which determines the execution rate. The frequency is defined by its reciprocal, that is, the period of time between two consecutive activations of the instance. For each of the four groups of experiments, we varied the period starting from 15ms and doubling it until 960ms. For example, a period of 15ms means that the feedback loop instance is executed every 15ms, which is only feasible if the total computation time of the feedback loop plus the overhead of the code-based solution or the \EUREMA interpreter is below 15ms. 

\begin{figure}[t]
\begin{minipage}[b]{0.48\linewidth}
\centerline{\includegraphics[height=45mm]{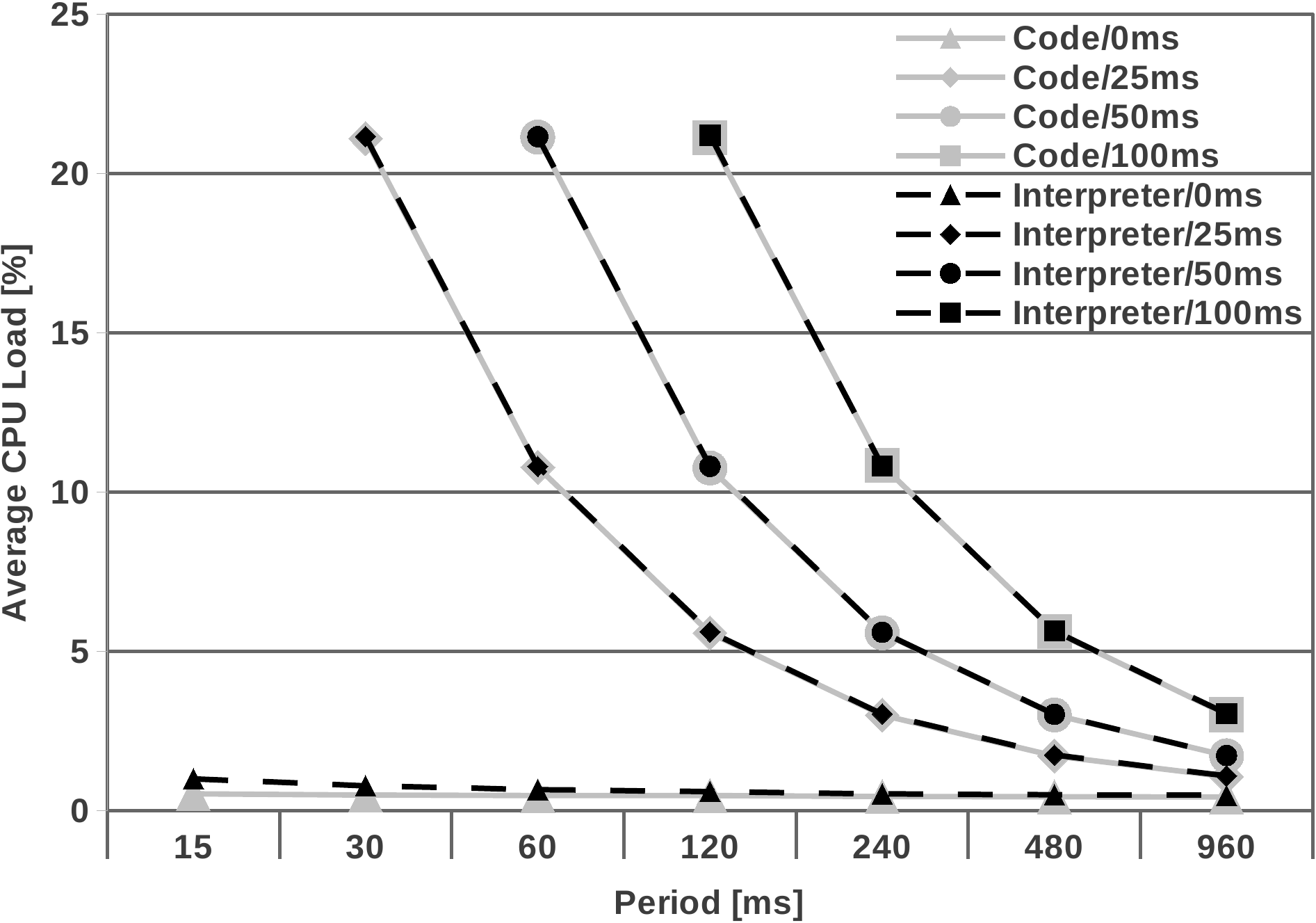}}
\caption{Average CPU Load of the code-based solution (\textbf{Code}) and \EUREMA (\textbf{Interpreter})}
\label{fig:chart:absolute-load}
\end{minipage}%
\hspace{0.02\linewidth}
\begin{minipage}[b]{0.48\linewidth}
\centerline{\includegraphics[height=45mm]{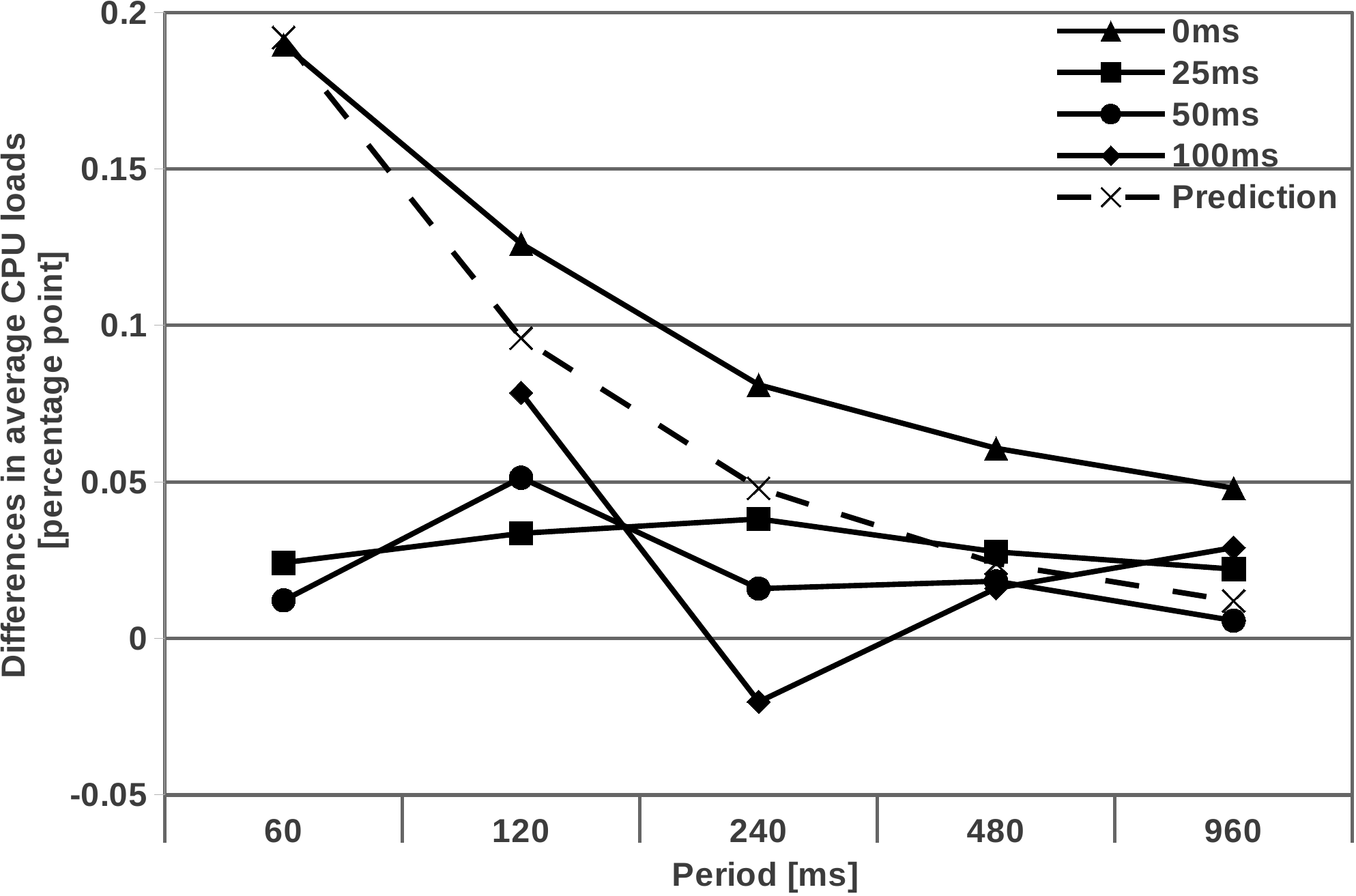}}
\caption{Interpreter overhead by means of the differences in average CPU loads}
\label{fig:chart:difference-load}
\end{minipage}
\end{figure}

For each feasible combination of these two parameters, we measured the load of the Java virtual machine for the code-based solution and the \EUREMA interpreter while~executing the self-repair feedback loop for a total time of ten minutes. 
The results of the experiments\footnote{~The experiments were conducted on the following platform: quad-core CPU (Intel Core i5-2400, 3.10GHz), 8GB RAM, Ubuntu 12.04 (Kernel 3.2.0-33), Java SE Runtime Environment 1.6.0\_31, and Eclipse Modeling Framework (EMF) Runtime and Tools 2.7.2. The CPU load has been measured by the monitoring capabilities of \emph{Java VisualVM} provided with the Java Development Kit 6 (1.6.0\_31).} are depicted in Figures~\ref{fig:chart:absolute-load} and~\ref{fig:chart:difference-load}.
Figure~\ref{fig:chart:absolute-load} visualizes the average CPU load of the code-based solution (solid gray lines) and the \EUREMA interpreter (dashed black lines) for the different frequencies of executing the feedback loop~instance. Moreover, each graph refers to a specific total computation time of the feedback loop instance (see legend).
Based on this figure, we may generally observe that the average load decreases for both solutions and all computation times of the feedback loop instance if the period between consecutive runs of the feedback loop \mbox{instance}~increases. This observation is not surprising since running a feedback loop less frequently is supposed to cause less load. 
Moreover, we may observe that the \EUREMA interpreter causes slightly more load than the code-based solution when the computation time of the feedback loop is 0ms. However, for the other cases of the computation time, there are no apparent differences between the loads of the code-based solution and the interpreter, and the corresponding graphs overlap. 
Thus, the overhead of the interpreter is noticeable for the hypothetical case that the feedback loop does not perform any computations and therefore, the computations do not cause any load. 

To further investigate the overhead, we calculated the overhead as the difference between the average loads of the interpreter and the average loads of the code-based solution for each case of the total computation time (cf. Figure~\ref{fig:chart:difference-load}). 
We may observe that for all cases the overhead of the \EUREMA interpreter with respect to the code-based solution is always below 0.2 percentage points and tends to decrease with increasing frequency periods. This assumptions is supported by the overhead we predicted (cf. \emph{Prediction} graph), which is the average overhead based on all measurements for all frequencies and computation times, and normalized for the frequencies.

Summing up, the experiments show that the overhead of interpreting \EUREMA models is negligible. In particular, the hypothetical case when the feedback loop's operations do not perform any internal computations revealed the pure load caused by the \EUREMA interpreter. The average of this pure load was for all experiments below 1\% (cf. \emph{Interpreter/0ms} graph in Figure~\ref{fig:chart:absolute-load}). Thus, in absolute terms, the \EUREMA interpreter works efficiently for the considered case of executing the \EUREMA models specifying the self-repair feedback loop. Moreover, employing the \EUREMA interpreter and accepting its overhead provides the flexibility to dynamically adapt feedback loops as discussed in Sections~\ref{sec:ld:layered-architecture} and~\ref{sec:ld:offline-adaptation}.

The validity of the experiments is threatened since we implemented the alternative, code-based solution, such that the comparison of this solution with the \EUREMA interpreter needs further investigations. Nevertheless, we have shown that the interpreter works efficiently in absolute terms by causing a negligible average load (cf. previous paragraph).
Another threat of validity is the specific self-repair feedback loop we used. However, we think that this feedback loop is a typical one since it follows the MAPE-K principle like the state-of-the-art approaches (Rainbow, DiVA, and PLASMA) \cite{VG-TR13}. Moreover, the complexity of the specific feedback loop by means of the~numbers of model operations and runtime models can be questioned and how the interpreter behaves for larger \EUREMA models. Thus, the scalability of the interpreter, which is determined by the specific \EUREMA models and not by the \EUREMA language, needs to be further investigated though state-of-the-art approaches (cf. \citeN{VG-TR13}) do not employ significantly more complex feedback loops with respect to the size of the \FeedbackLoopDiagrams.

\section{Conclusion and Future Work}\label{sec:conclusion}

In this article, we presented \EUREMA, a model-driven approach for engineering adaptation engines for self-adaptive software. \EUREMA provides a domain-specific modeling language to specify and an interpreter to execute feedback loops.  
In contrast to existing work on self-adaptive software, \EUREMA is a seamless approach that covers the specification and the execution of adaptation engines. Related approaches on modeling languages provide no runtime support for their models while related work on frameworks does not support the explicit modeling of feedback loops. Moreover, frameworks do not provide the flexibility for their users in structuring user-defined adaptation activities to form arbitrary  feedback loops in an arbitrary number of layers. In this context, \EUREMA does not impose any restriction.

The \EUREMA language supports the explicit modeling of feedback loops and their coordinated execution. Thereby, the runtime models as the feedback loop's knowledge are explicitly captured. Moreover, \EUREMA models are kept alive at runtime, which leverages layers of feedback loops for dynamically adjusting the adaptation engine as well as evolving the self-adaptive software by offline adaptation. Therefore, this article has discussed \EUREMA with respect to the requirements reflecting state-of-the-art problems for self-adaptive software and \EUREMA provides concepts for almost all of them. 
This demonstrates that \EUREMA is expressive enough to cover a wide range of problems for self-adaptive software at the conceptual level. 
To complete the evaluation, we further investigated the expressiveness by examples. Therefore, we modeled three state-of-the-art approaches from the literature. Finally, we evaluated the \EUREMA interpreter by quantifying its runtime characteristics, which shows evidence that the interpreter works efficiently and is applicable at runtime.

As future work, we plan to further elaborate \EUREMA by considering the requirements for self-adaptive software we are currently not addressing. Thus, we want to investigate the concurrent execution of interdependent feedback loops and the distribution of adaptation engines in the context of a multi-robot system. 
Furthermore, we want to investigate the integration of model-based techniques to analyze and simulate modular feedback loop specifications. 
Finally, we want to further evaluate \EUREMA by conducting empirical studies.

\vspace{1em}
\bibliographystyle{acmsmall}
\bibliography{references}

\received{November 2012}{January 2013, July 2013, November 2013}{November 2013}
\end{document}